\documentclass[a4paper,11pt]{article}
\pdfoutput=1
\usepackage{graphicx,rotating,hyperref,slashed,amsmath,xcolor,amssymb,amsfonts,colortbl,cite, subfigure,float}
\makeatletter 
\usepackage{graphics}
\usepackage{graphicx} 
\graphicspath{{plots/}}
\numberwithin{equation}{section}

\hypersetup{colorlinks,bookmarksopen,bookmarksnumbered,
linkcolor=blus,pdfstartview=FitH,urlcolor=rossos,citecolor=verde}

\hypersetup{%
    ,urlcolor=black
    ,citecolor=black
    ,linkcolor=black
    }

\AtBeginDocument{
  \hypersetup{
    urlcolor=blue,
    citecolor=blue,
    linkcolor=black,
  }%
}

\allowdisplaybreaks

\def\lsim{\mathrel{\rlap{\lower3pt\hbox{\hskip0pt$\sim$}}
   \raise1pt\hbox{$<$}}}         
\def\gsim{\mathrel{\rlap{\lower4pt\hbox{\hskip1pt$\sim$}}
   \raise1pt\hbox{$>$}}}         

 \newcommand{\sfootnote}[1]{} 
\definecolor{bluc}{cmyk}{1,1,0,0.1}
\definecolor{rossoCP3}{cmyk}{0,.88,.77,.40}
\definecolor{rosso}{cmyk}{0,1,1,0.4}
\definecolor{rossos}{cmyk}{0,1,1,0.55}
\definecolor{rossoc}{cmyk}{0,1,1,0.2}
\definecolor{verdes}{cmyk}{0.92,0,0.59,0.4}

\hypersetup{colorlinks, bookmarksopen, bookmarksnumbered,
citecolor=verdes, linkcolor=bluc, pdfstartview=FitH, urlcolor=rossos}

\newcommand{\mio}[1]{}

\definecolor{Gray}{gray}{0.95}

\usepackage{multicol}
\usepackage{color}
\definecolor{rosso}{cmyk}{0,1,1,0.4}
\definecolor{rossos}{cmyk}{0,1,1,0.55}
\definecolor{rossoc}{cmyk}{0,1,1,0.2}
\definecolor{blu}{cmyk}{1,1,0,0.3}
\definecolor{blus}{cmyk}{1,1,0,0.6}
\definecolor{bluc}{cmyk}{1,1,0,0.1}
\definecolor{verde}{cmyk}{0.92,0,0.59,0.25}
\definecolor{verdec}{cmyk}{0.92,0,0.59,0.15}
\definecolor{verdes}{cmyk}{0.92,0,0.59,0.4}

\def\circa#1{\,\raise.3ex\hbox{$#1$\kern-.75em\lower1ex\hbox{$\sim$}}\,}

\newcommand{\beq}{\begin{equation}}
\newcommand{\eeq}{\end{equation}}

\newcommand{\bea}{\begin{eqnarray}}
\newcommand{\eea}{\end{eqnarray}}
\newcommand{\be}{\begin{equation}}
\newcommand{\ee}{\end{equation}}
\newfam\rsfsfam
\def\mathscr#1{{\fam\rsfsfam\relax#1}}

\def\circa#1{\,\raise.3ex\hbox{$#1$\kern-.75em\lower1ex\hbox{$\sim$}}\,}
\makeatletter

\def\hhref#1{\href{http://arxiv.org/abs/#1}{arXiv:#1}} 

\newcommand{\doi}[1]{\href{http://dx.doi.org/#1}{[doi]}}

\def\hhref#1{\href{http://arxiv.org/abs/#1}{arXiv:#1}} 
 
\def\art{\@ifnextchar[{\eart}{\oart}}
\def\eart[#1]#2#3#4#5#6{{\rm #2}, {\em #3 \bf #4} {\rm (#6) #5} ({\em #1})}

\def\article{\@ifnextchar[{\earticle}{\oarticle}}
\def\oarticle#1#2#3#4#5#6{{\rm #1}, {\em ``#6''}, {\rm #2 #3 (#5) #4}}
\def\earticle[#1]#2#3#4#5#6#7{{\rm #2}, {\em ``#7''}, {\rm #3 #4 (#6) #5}  [\hhref{#1}]}
\def\hepart[#1]#2{{\rm #2, \em#1}}
\def\heparticle[#1]#2#3{#2, {\em ``#3''} [\hhref{#1}]}

\newcounter{alphaequation}[equation]
\def\thealphaequation{\theequation\hbox to
0.6em{\hfil\alph{alphaequation}\hfil}}
\def\eqnsystem#1{
\def\@eqnnum{{\rm (\thealphaequation)}}
\def\@@eqncr{\let\@tempa\relax \ifcase\@eqcnt \def\@tempa{& & &} \or
  \def\@tempa{& &}\or \def\@tempa{&}\fi\@tempa
  \if@eqnsw\@eqnnum\refstepcounter{alphaequation}\fi
\global\@eqnswtrue\global\@eqcnt=0\cr}
\refstepcounter{equation} \let\@currentlabel\theequation \def\@tempb{#1}
\ifx\@tempb\empty\else\label{#1}\fi
\refstepcounter{alphaequation}
\let\@currentlabel\thealphaequation
\global\@eqnswtrue\global\@eqcnt=0 \tabskip\@centering\let\\=\@eqncr
$$\halign to \displaywidth\bgroup \@eqnsel\hskip\@centering
$\displaystyle\tabskip\z@{##}$&\global\@eqcnt\@ne
\hskip2\arraycolsep\hfil${##}$\hfil& \global\@eqcnt\tw@\hskip2\arraycolsep
$\displaystyle\tabskip\z@{##}$\hfil
\tabskip\@centering&\llap{##}\tabskip\z@\cr}
\def\endeqnsystem{\@@eqncr\egroup$$\global\@ignoretrue} \makeatother

\definecolor{fiorentina}{rgb}{.5,0,.5}

\begin{document}

\vspace{1truecm}
 \begin{center}
\boldmath

{\textbf{\Large
Gravitational wave non-linearities  and 
  pulsar-timing array\\ \vspace{0.1cm}
  angular correlations 
}}
\unboldmath
\unboldmath
\hskip1cm

\end{center}

\vspace{0.2cm}

\begin{center}
{\fontsize{13}{30}\selectfont  Gianmassimo Tasinato } 
\end{center}

\begin{center}

\vskip 8pt
\textsl{ Physics Department, Swansea University, SA28PP, United Kingdom }\\
\vskip 7pt

\end{center}

\vspace{1.2cm}
 \vspace{0.3cm}
\begin{abstract}
\noindent
Several pulsar-timing array (PTA) collaborations are finding tantalising  hints for  a stochastic gravitational wave background  signal in the nano-Hertz regime. So far, though, no  convincing evidence for  the expected Hellings-Downs quadrupolar  correlations has been found. While this issue might  get fixed  at the light of more accurate, forthcoming  data, it is   important to keep an  eye open on  different possibilities, and  explore scenarios able to   produce different types of  PTA angular correlations. We point  out that a  stationary  non-Gaussian component to  the gravitational wave background can modulate the 2-point PTA overlap reduction function, adding contributions that can  help in fitting  the angular distribution of   PTA data. We discuss possible  sources for such non-Gaussian signal in terms of cosmological processes occurring after inflation ends, and we    investigate  further    tests for this idea.
\end{abstract}

\vskip1cm

\section{Introduction}
\label{sec_theory}

Pulsar-timing arrays (PTA) offer a promising tool for detecting gravitational waves (GW) in the nano-Hertz regime. The concept was first proposed in \cite{Sazhin:OO,Detweiler:1979wn,Mashhoon:1979wk,Bertotti:1980pg}, and much developed thereafter  -- see e.g. \cite{Lommen:2015gbz} for a  review. 
Recently, the NANOGrav collaboration detected a signal compatible with a stochastic gravitational wave background (SGWB)  \cite{NANOGrav:2020bcs}. Subsequently, the PPTA \cite{Goncharov:2021oub},
 EPTA \cite{Chen:2021rqp}  and IPTA \cite{Antoniadis:2022pcn} collaborations obtained preliminary results going in the same direction. 
A natural 
astrophysical source for such a SGWB  is constituted by unresolved GW signals from super-massive black hole mergers \cite{Haehnelt:1994wt,Sesana:2004sp,Sesana:2008mz}. However,  PTA GW detections can also be explained by   cosmological sources as GW echoes from primordial
black hole formation \cite{Vaskonen:2020lbd,DeLuca:2020agl,Kohri:2020qqd}, cosmic strings \cite{Ellis:2020ena,Buchmuller:2020lbh,Blasi:2020mfx,Blanco-Pillado:2021ygr},  phase transitions \cite{Nakai:2020oit,Ratzinger:2020koh,Addazi:2020zcj,NANOGrav:2021flc,Brandenburg:2021tmp},  or 
primordial magnetic field production \cite{Neronov:2020qrl,RoperPol:2022iel}. A puzzling
feature of  PTA measurements  so far  is that 
the constraints on spatial correlations seem to  show  some  deviations from
the Hellings-Downs (HD) quadrupolar  
%
%
angular distribution  \cite{Hellings:1983fr}, which is a consequence of   Einstein General Relativity. See e.g. Fig 7 in  \cite{NANOGrav:2020bcs} for NANOGrav;
Fig  3 in \cite{Goncharov:2021oub} for PPTA; Fig 2 in  \cite{Chen:2021rqp} for EPTA.
  In
case  anomalous angular correlations  are present 
  after more data are collected,
 %
 they  will require some departure from the  
standard approach.  A possibility, considered for example in \cite{Chen:2021wdo}, is that
    that   NANOGrav   is detecting   extra GW polarizations besides 
 Einstein's spin-2 ones  \cite{Eardley:1973zuo,Eardley:1973br}, since the inclusion of non-Einsteinian  polarizations modifies the  HD angular distribution \cite{Chamberlin:2011ev,Gair:2015hra,Cornish:2017oic,Romano:2016dpx}. A systematic analysis by the NANOGrav collaboration does not presently favour
 this option \cite{NANOGrav:2021ini}, but it  recommends to study this topic further
  at the light of  
 forthcoming data. In this context, however,  we point out  that a  recent analysis from the LIGO-Virgo-Kagra collaboration
 does not provide  evidence for non-Einsteinian GW polarizations in the deci-Hertz regime \cite{LIGOScientific:2021sio}.
 
 \smallskip
 
 In this work, motivated from the  aforementioned  preliminary results of PTA observations,  we 
 %
   explore   
  an alternative mechanism  
  for  modifying the quadrupolar HD    angular distribution, using only the massless spin-2  degrees of freedom of General Relativity. 
   We show that large tensor non-Gaussianity 
   can modulate  the angular 2-point PTA overlap reduction function (ORF), and parametrically change its  profile as a function of the angle between pulsars. 
   %
The non-Gaussianity of the SGWB is an observable not often considered in the GW literature.   It has been realized since long time that, 
   thanks to the central limit theorem, astrophysical SGWB signals are expected to be Gaussian, 
   being the cumulative contribution
    of many unresolved sources \cite{Allen:1987bk}. However, cosmological SGWB sources, 
     -- inflation, phase transitions etc --    are coherent and can be  characterized by 
     large tensor non-Gaussianity -- see e.g. \cite{Bartolo:2018qqn}, section 5 for a review. 
   In general, tensor non-Gaussianity from cosmological sources  can {\it not} be directly measured 
  with GW experiments, since it leads to  non-stationary signals which lose their crucial phase
  correlations in their way from emission to detection \cite{Bartolo:2018evs,Bartolo:2018rku,Margalit:2020sxp} \footnote{Similar effects were  previously studied in \cite{Allen:1999xw} in the context of 2-point functions from inflation.  Notice that   indirect effects of tensor non-Gaussianities can be detected   through  correlators
  of SGWB anisotropies \cite{Bartolo:2019oiq,Bartolo:2019yeu,Dimastrogiovanni:2019bfl}.}. A possible way out is to focus on the specific momentum shape corresponding to folded tensor non-Gaussianities, that in real space leads to a stationary signal 
  that  does not necessarily suffer from the aforementioned problems \cite{Powell:2019kid}. Folded
  non-Gaussianities can arise  in scenarios where  a stochastic background is generated by   {\it causal}  sources \cite{Green:2020whw}  \footnote{The work \cite{Green:2020whw} specifically focusses  on scalar fluctuations, but its  general arguments
  apply to tensor fluctuations as well. (See also appendix \ref{app_model}.)}. 
     In fact, as described in  \cite{Green:2020whw}, a folded non-Gaussian  shape
  is  associated
  with poles at physical momenta in the connected $n$-point functions, and is a
   consequence of non-Gaussian cosmological signals produced by mechanisms
  that preserve locality and causality. 
   Explicit computations of tensor non-Gaussianities from post-inflationary  cosmological  sources are carried out  in \cite{Adshead:2009bz}, including
   scenarios of cosmological   phase transitions  capable to generate
    connected  $n$-point correlators ($n>2$), 
   with an amplitude  comparable to  the one of
  $2$-point correlators.
   The work \cite{Adshead:2009bz} focussed on equilateral configurations for $n$-point functions in momentum space though, without discussing  folded configurations that  -- as argued in   in \cite{Green:2020whw} -- can   generally  contribute to classical $n$-point functions in Fourier space.
  
\smallskip

Our discussion proceeds as follows. In section \ref{sec_two_point} we show how non-linear effects associated with tensor
non-Gaussianity can modulate the angular distribution 2-point overlap reduction function. 
 We compute how the resulting ORF profile 
depends on quantities characterizing the higher-order tensor correlation functions. In 
the hypothesis that  next releases of 
 PTA data will show  a  significant departure from  HD  angular correlations, it will be important to design
tests to distinguish among different explanations for this phenomenon. For this reason, 
as a specific prediction of the proposal elaborated in section \ref{sec_two_point}, in
section \ref{sec_four_point} we analyse   $4$-point  connected correlation functions
of PTA signals, showing that their detection would indicate  the presence of tensor non-Gaussianities in the SGWB.  Section \ref{sec_conclusions} contains our conclusions, which are followed by four technical appendixes, elaborating
the results presented in the main text.

\smallskip

\bigskip



\section{Modulation of the PTA 2-point overlap  reduction function}
\label{sec_two_point}

In this section we show that stationary tensor non-Gaussianity can affect
the 2-point function of PTA signals, and parametrically change the angular distribution of the
corresponding overlap reduction function (ORF), with respect to the Hellings-Downs (HD)
curve. In this work we take a phenomenological perspective, leaving a more
general treatment and  a systematic investigation of  model building to future studies.

\smallskip

In order to describe a SGWB, 
we express the GW modes   in terms
of 
 spin-2 
fluctuations around  flat space as
\be\label{expanze}
g_{\mu\nu}\,d x^\mu d x^\nu\,=\,-d t^2+\left(\delta_{ij}+h_{ij}(t,\vec x) \right) \,d x^i d x^j\,,
\ee
 with
$h_{ij}(t, \vec x)$  the  tensor fluctuation satisfying the transverse-traceless condition $h_{\,\,i}^i\,=\,\partial^i\,h_{ij}\,=\,0$.

The presence of a GW deforms  light geodesics, and induces a time delay $\Delta T_\alpha$ on the period $T_\alpha$
of a pulsar $\alpha$, located at a position $\vec x_\alpha\,=\,\tau_\alpha\,\hat x_\alpha$ with respect to  the Earth at $\vec x=0$. (We denote  $\tau_\alpha$ the travel time from source to detection, and we set $c=1$ from now on.) Denoting with  $\hat n$ the direction
of the GW,  and  introducing the convenient combination
\be
E_{\alpha}(t,\vec x)\,\equiv\,\hat x^i_\alpha \hat x^j_\alpha \,h_{ij}(t,\vec x)\,,
\ee
 we find the following expression for the time delay $z_\alpha$ induced by the GW:
\bea
z_\alpha&\equiv&
\frac{\Delta T_\alpha(t)}{T_\alpha}
\,=\,
\frac{1}{2(1+\hat x_\alpha\cdot \hat n)}
\Big[E_{\alpha}(t,\vec x=0)- E_{\alpha}(t-\tau_\alpha,\vec x=\vec x_\alpha) 
\nonumber
\\
&&\hskip4.5cm -\frac34 \left( 
E^2_{\alpha}(t,\vec x=0)- E^2_{\alpha}(t-\tau_\alpha,\vec x=\vec x_\alpha) 
\right)
\nonumber
\\
&&\hskip4.5cm +\frac{5}{16} \left( 
E^3_{\alpha}(t,\vec x=0)- E^3_{\alpha}(t-\tau_\alpha,\vec x=\vec x_\alpha) 
\right)+\dots
\Big]
\label{TaGEN1}\,.
\eea
The first line of eq \eqref{TaGEN1} is the classic result  of  \cite{Detweiler:1979wn}
(see e.g. \cite{Maggiore:2018sht}, Chapter 23 for a textbook derivation).
 The second and
third lines are higher order  corrections associated with non-linearities in $h_{ij}$, and
are a new result  of this work.
We present in appendix \ref{appTDPTA} a derivation of eq \eqref{TaGEN1}, including a generalization
valid for any power in an expansion in   $E_{\alpha}$.   

\smallskip

We proceed analyzing  here  the  2-point correlation
functions among pulsar time delays, given by $\langle z_\alpha z_\beta\rangle
$ (with $\alpha$, $\beta$ denoting the two pulsars). 
Given the non-linear structure of eq \eqref{TaGEN1}, we  expect that the PTA 2-point correlator
is  modulated  by   higher order connected  $n$-point  functions involving spin-2 fluctuations $h_{ij}$. 
This phenomenon can change the angular dependence of the PTA overlap function. 

\smallskip

 We expand the transverse-traceless GW gauge
in Fourier modes as
\be 
\label{sig_fourier}
h_{ij} (t,\vec x)\,=\,
\sum_\lambda \int_{-\infty}^\infty  d f \int d^2 \hat n\,e^{-2\pi\,i\,f\,\hat n\,\vec x }
\,
e^{2\pi\,i\,f\,t}\,
{\bf e}_{ij}^{(\lambda)}(\hat n)\,
 h_{\lambda}(f,\,\hat n)\,,
\ee
where $f$ is the GW frequency, and the unit vector $\hat n$ controls its direction.   We formally integrate over positive as well as negative frequencies, and
the reality of $h_{ij}(t,\vec x)$ imposes the condition $h_\lambda(-f,\,\hat n)\,=\,h_\lambda^*(f,\hat n)$ on the Fourier modes. The polarization states for the spin-2 fields are $\lambda\,=\,(+, \times)$.  Our conventions for the polarization tensors ${\bf e}_{ij}^{(\lambda)}(\hat n)$, and
some of their properties, are spelled out in appendix \ref{app_A}. 

\begin{figure}[h!]
\centering
  \includegraphics[width = 0.5 \textwidth]{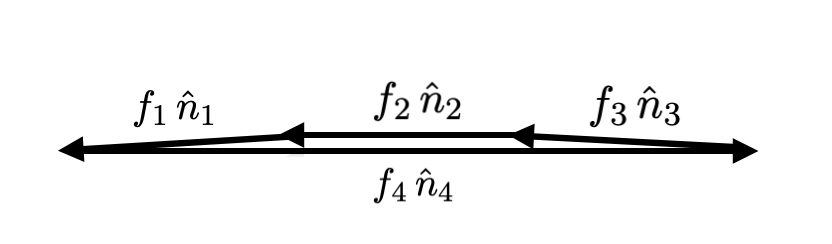}
 \caption{\small A folded quadrangle configuration for momenta in Fourier space, satisfying
 the $\delta$-function conditions of eq \eqref{def4pfFa}. The short sides of the quadrangle
 are superimposed on the long one.}
 \label{fig:plot0a}
\end{figure}

We assume that the GW spectrum is unpolarized, with a 2-point function given by 
\bea
\label{ans2ptF}
\langle  h_{\lambda_1}(f_1,\,\hat n_1)\, h_{\lambda_2}(f_2,\,\hat n_2) \rangle\,=\,\delta_{\lambda_1\lambda_2}\,\delta(f_1+f_2)\,\delta^{(2)}(\hat n_1-\hat n_2) \,P(f_1)\,,
\eea
where the $\delta$-function conditions are associated with momentum conservation.
 Additionally, we assume that the SGWB is non-Gaussian, and we parameterise its  properties 
  in terms of
   a non-vanishing 4-point function  in momentum space:
\bea
\langle \, \Pi_{i=1}^4\,h_{\lambda_i}(f_i,\,\hat n_i)
  \rangle&=&  \Pi_{i=1}^3\,\delta^{(2)}(\hat n_4-\hat n_i)\,\delta(f_4+3 f_i)
\,\times\,
H_{\lambda_1 \dots 
\lambda_4}(f_4)\,
 { P}(f_4)
 \label{def4pfFa}\,.
 \eea
The non-Gaussian shape associated with eqs \eqref{def4pfFa} 
 corresponds to a folded quadrangle, with three  small  sides
of the quadrangle of equal length and superimposed on the fourth,  longest one (see Fig \ref{fig:plot0a}).   We focus on the 4-point function as \cite{Seto:2009ju}, being
a  convenient quantity  in treating the modulation effects of the ORF \footnote{  In fact,  applying the procedure  of \cite{Seto:2009ju} to the PTA case, one finds that contributions
from the 3-point function vanish, while  the 4-point function is able to modulate the ORF: see  appendix
\ref{app_A}.}.
The condition that
the short length sides of the folded quadrangle are equal -- as forced by the $\delta(f_4+3 f_i)$ conditions in \eqref{def4pfFa} --  is chosen for simplifying our arguments.  The amplitude
in \eqref{def4pfFa} is proportional to the power spectrum $P(f)$ as introduced 
in eq \eqref{ans2ptF}.
 We include as coefficient of eq \eqref{def4pfFa}  a model-dependent tensor
$H_{\lambda_1\dots \lambda_4}(f)$, depending on the polarization indexes and on frequency.
A folded tensor non-Gaussianity 
can arise in scenarios where a cosmological SGWB is produced by
cosmological sources  after inflation ends. In fact, interactions in such scenarios  preserve locality and 
causality, and lead 
to characteristic poles in higher-order correlation functions which  amplify non-Gaussian
folded shapes. We refer to
 appendix 
\ref{app_model} for additional explanations and an explicit example.

\noindent
A folded
non-Gaussianity in Fourier space 
 leads
to a stationary 4-point function in real space \cite{Powell:2019kid}:
\bea
\langle \Pi_{m=1}^4\,h_{i_m j_m}(t_m, \,\vec x_m) \rangle&=&\sum_{\lambda_i}
\int d f\,d^2 \hat n\,e^{2 \pi i f\, \left[  (t_1-t_4)+ (t_2-t_4)+ (t_3-t_4)\right]}\,
e^{-2 \pi i\,f\,\hat n\,\left[  (\vec x_1-\vec x_4)+ (\vec x_2-\vec x_4)+ (\vec x_3-\vec x_4) \right]}
\nonumber\\
&&\,\,\,
\,\,\,P(f)\,\left[H_{\lambda_1  \lambda_2 \lambda_3 
\lambda_4}(f) \,{\bf e}_{i_1 j_1}^{(\lambda_1)}(\hat n)\,{\bf e}_{i_2 j_2}^{(\lambda_2)}(\hat n)\,{\bf e}_{i_3 j_3}^{(\lambda_3)}(\hat n)\,{\bf e}_{i_4 j_4}^{(\lambda_4)}(\hat n)
\right]
\label{statc1}
\,,
\eea
which depends on time and on space differences only.  
As a matter of principle, the stationarity
condition in eq  \eqref{statc1} can allow us  to circumvent  the arguments developed in  \cite{Bartolo:2018evs,Bartolo:2018rku,Margalit:2020sxp}, which finds that tensor non-Gaussianity  can not  be directly measured with GW experiments:  Along  their  
way from source to detection, GW lose their phase correlations due to random effects associated with  Shapiro time delays induced by cosmic fluctuations. However, if measurements depend
on time differences only -- as in the stationary case of eq \eqref{statc1} -- cumulative disturbances  cancel out, and the
results depend only on the relatively small time-differences between successive measurements of
pulsar timing periods.
   In fact, we can assume that
any further (non-folded) contribution to the 4-point function leads to a non-stationary signal which   is not directly measurable
in terms of correlators of PTA measurements, and we focus on non-Gaussian contributions associated with  eq \eqref{def4pfFa} only.  See \cite{Powell:2019kid} for more details, and \cite{Allen:1999xw} for similar considerations
for the case of (non-)stationary contributions to primordial  2-point functions.

\smallskip

As mentioned above, we are assuming 
that the amplitude of the 4-point function in Fourier space, eq \eqref{def4pfFa},  is proportional to the power
spectrum $P(f)$ (times the model-dependent function of frequency and polarization
indexes,  $H_{\lambda_1\dots \lambda_4}$).
 Coherent cosmological  sources, which are able to amplify the GW spectrum by causal mechanisms,  
make use of strong non-linear interactions   for the fields involved. They are
expected to  enhance
not only the 2-point,  but also the $n$-point GW correlation functions, with $n>2$: 
   the amplitude of $n$-point correlators  can be of the same order of the $2$-point one 
 \cite{Adshead:2009bz}.  It would be interesting to study more systematically at what
 extent these phenomena  enhance folded limits of $n$-point correlation functions,
 depending on the scenarios considered. We discuss a preliminary example
 in Appendix \ref{app_model}, leaving more detailed analysis to future studies. 


 The  quantity $H_{\lambda_1 \dots 
\lambda_4}$  describes the dependence of the $4$-point function
on the helicity indexes $\lambda_i$. We phenomenologically parametrize it   as follows
\be
\label{ans4ptFb}
H_{\lambda_1  \lambda_2 \lambda_3 
\lambda_4}\,\equiv\,\kappa_1(f)\,\delta_{\lambda_1\lambda_2}\,\delta_{\lambda_3\lambda_4}
+\kappa_2(f)\,\left(1-\delta_{\lambda_1\lambda_2}\right)\,\left(1-\delta_{\lambda_3\lambda_4}\right)\,\delta_{\lambda_1\lambda_3}\,\delta_{\lambda_2\lambda_4}\,,
\ee
in terms of two frequency-dependent parameters $\kappa_{1,2}(f)$.
  It is straightforward   to consider more general forms for the tensor $H_{\lambda_1 \dots 
\lambda_4}$ as a function of the polarization indexes; we explored   other choices
and  found that the previous Ansatz  describes  well   the possible  angular dependences
of the 2-point overlap reduction functions. We stress that our hypothesis  are phenomenologically motivated by the aim  of making our considerations as transparent as possible.
They can be  generalised to study  more general cases,  as indicated 
by   specific model building.  

\bigskip

We make use of the results so far for computing the  equal time 2-point correlation functions
of two pulsar time delays, using formula eq \eqref{TaGEN1}. For the case of a  single GW propagating
through the direction $\hat n$, we find
\bea
\label{mod2pt}
\langle z_\alpha z_\beta\rangle &\equiv&
\langle \frac{\Delta T_\alpha}{T_\alpha}
 \frac{\Delta T_\beta}{T_\beta}
\rangle\,=\,\frac{ \langle E_{\alpha} E_{\beta}\rangle
+9/16 \left(  \langle E_{\alpha}^2 E_{\beta}^2\rangle\right)
+5/4 \left( \langle E_{\alpha}^3 E_{\beta}\rangle+ \langle E_{\alpha} E_{\beta}^3\rangle \right)
}{4 (1+\hat x_\alpha\cdot \hat n)(1+\hat x_\beta\cdot \hat n)}\,,
\eea
where the quantities  in this expression
are evaluated at the earth position $E_{\alpha,\beta}\,=\,E_{\alpha,\beta}(t,\,\vec x=0)$.
 The second and third term in the numerator of eq \eqref{mod2pt}
are new parts -- absent in the Gaussian case -- being   associated with  the higher order contributions in \eqref{TaGEN1}. These terms can modulate the overlap reduction functions, as we are going to learn. 
Contributions of `pulsar terms' of $E_{\alpha,\beta}$  to the 2-point functions, which
are  evaluated at  pulsar positions, are uncorrelated with the earth terms at $\vec x=0$.
They 
 lead to rapidly oscillating
pieces  when integrating over frequencies, 
   and can be neglected in the present instance  as in the standard Gaussian case (see e.g. the discussion in  \cite{Maggiore:2018sht}).

\begin{figure}[h!!]
\centering
  \includegraphics[width = 0.6 \textwidth]{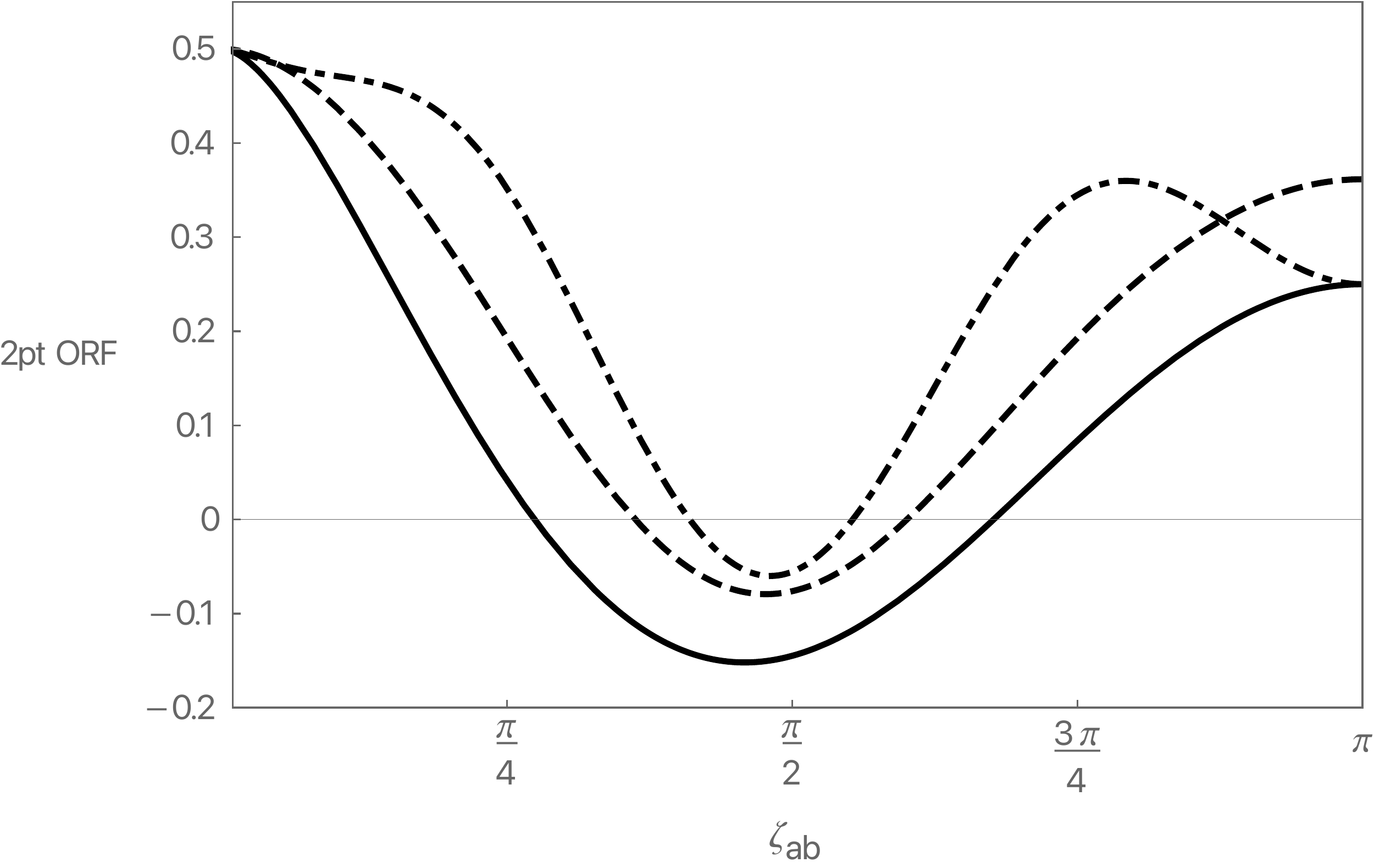}
 \caption{\small  The PTA 2-point overlap reduction function of eq \eqref{res2pt1}. Solid
 line: $\kappa_1=\kappa_2=0$ in eq \eqref{res2ptOV1} (HD curve).  Dot-dashed
 line:  $\kappa_1=2$, $\kappa_2=-8$ in eq \eqref{res2ptOV1}. Dashed line:  $\kappa_1=10$, $\kappa_2=0$ in eq \eqref{genorf2a} (we  normalize  the curve in such a way that its value matches 1/2 at $\zeta_{\alpha\beta}=0$, as for the HD curve, see the explanation after eq \eqref{genorf2a}).}
 \label{fig:plot1a}
\end{figure}

 We explicitly carry on  the calculation of the 2-point correlator $\langle z_\alpha z_\beta\rangle$ in appendix \ref{app_A}: the result can be expressed as
\bea
\langle z_\alpha z_\beta\rangle
&=&
\frac{8\pi}{3}\int d f\,P(f)\,\Gamma_{\alpha \beta}(f)
\,.
\label{res2pt1}
\eea
The overlap reduction function $\Gamma_{\alpha \beta}(f)$, for the case $\kappa_2\,=\,-4 \kappa_1$, results
\bea
\Gamma_{\alpha \beta}(f)&=&\frac12-\frac{x_{\alpha\beta}}{4} \left(1-\frac{54}{5} \kappa_1 \right)-\frac{171\, \kappa_1\,x_{\alpha\beta}^2}{10}+\frac{72  \kappa_1\, x_{\alpha\beta}^3}{5}+\frac32\,x_{\alpha\beta}\,\left( 1-9\,\kappa_1 \,x_{\alpha\beta}^2\right)\,\ln{x_{\alpha\beta}}\,,
\nonumber
\\
\label{res2ptOV1}
\eea
with 
\be
\label{defxab}
x_{\alpha\beta}\,\equiv\,\frac12\,(1-\cos \zeta_{\alpha\beta})\,,
\ee
 and $ \zeta_{\alpha\beta}$ 
the angle between the two  vectors controlling the pulsar positions $\vec x_\alpha$, $\vec x_\beta$ with respect to the earth.  
 We understand the dependence on frequency of $\kappa_1$, and the more general case of arbitrary $\kappa_{1,2}$ is discussed in appendix \ref{app_A}.  Notice that when $\kappa_1\,=\,\kappa_2\,=\,0$ we recover
 the standard HD curve.  We plot the corresponding
2-point ORF in Fig \ref{fig:plot1a} for some representative choices of constant parameters $\kappa_{1,2}$. 

\smallskip

The new  ORF
profiles shown in  Fig \ref{fig:plot1a} have the tendency to smooth the  anticorrelations characterizing  the HD curve for
angular separations $\zeta_{\alpha\beta}\simeq \pi/2$. This reduction of anticorrelations is  a feature in common with other ORF profiles,
as the ones induced by a  monopole, or a scalar contribution (see e.g.
   \cite{NANOGrav:2021ini}).
 In the present instance, we refrain from pursuing a proper fit of our
parametrization \eqref{res2ptOV1} with existing PTA data, and from performing a dedicated statistical analysis. In fact, current results  still have
systematic uncertainties (for example, in modelling  Solar System ephemeris, as explained in \cite{NANOGrav:2021ini}) which will be cured by more accurate, 
forthcoming data releases. But above all,  in our scenario the  time-residuals correlators  are  non-Gaussian, hence {we can
not use} the  statistical  methods based on Gaussian multidimensional likelihoods (see e.g. \cite{Anholm:2008wy,vanHaasteren:2008yh}). We should  elaborate  a dedicated
analysis to the  non-Gaussian context we are interested in, also  including the frequency dependence
for the quantities $\kappa_{1,2}$ appearing in our ORF as motivated by specific models.   This goes beyond the scope of this theoretical work, and we postpone it to future analysis \footnote{But see for example \cite{Lentati:2014hja} for interesting attempts to include non-Gaussian statistics in the modelling uncorrelated noise sources affecting  PTA GW detections.},  in the  case that forthcoming  
PTA data will not favour HD-type angular correlations.

\begin{figure}[h!!]
\centering
  \includegraphics[width = 0.5 \textwidth]{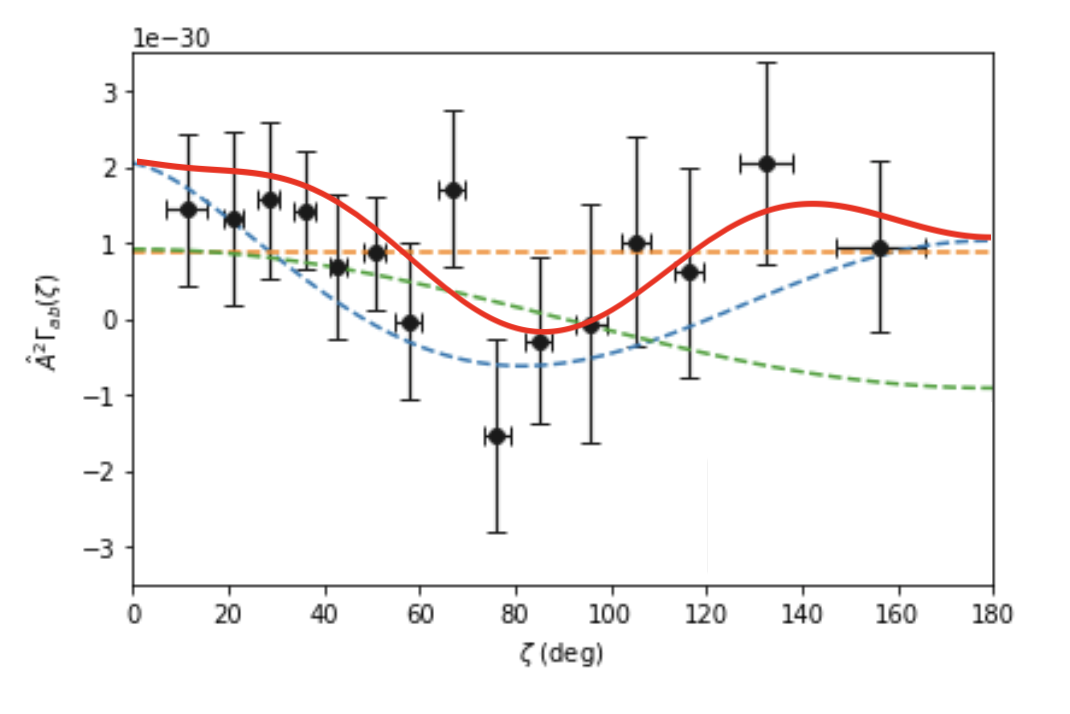}
 \caption{\small Visual representation of how the modulated ORF 
of eq  \eqref{res2ptOV1} with   $\kappa_1=2$, $\kappa_2=-8$, shown as continuous red line,  compares
against binned NANOGrav 12.5 data \cite{NANOGrav:2020bcs}. We include additional ORF profiles, as \cite{NANOGrav:2020bcs}:  Dashed blue: Hellings-Down curve; Dashed orange: monopole ORF; Dashed green: dipole ORF. Figure obtained starting from the content  in    \href{https://data.nanograv.org/}{{\ttfamily
 https://data.nanograv.org/}}. To allow for a more direct visual  comparison, we use  the same conventions and notation of \cite{NANOGrav:2020bcs} (see their Fig 5).
}
 \label{fig:plot4a}
\end{figure}

\smallskip

Nevertheless, for visual aid only, we represent in Fig \ref{fig:plot4a} in red colour our ORF profile of eq \eqref{res2ptOV1} against  binned NANOGrav 12.5 data \cite{NANOGrav:2020bcs}, choosing
  $\kappa_1=2$, $\kappa_2=-8$. We also include
the HD,  the monopole, and dipole ORF profiles. This figure is obtained starting from the content  in the NANOGrav tutorial website 
 \href{https://data.nanograv.org/}{{\ttfamily
 https://data.nanograv.org/}},  based on \cite{Anholm:2008wy,Demorest:2012bv,Chamberlin:2014ria,Vigeland:2018ipb},  which explains how to obtain Fig 5 of \cite{NANOGrav:2020bcs}.   Suggestively, the red line correponding to our ORF apparently fits the data well.


\section{ A test: 4-point correlation functions of PTA signals}
\label{sec_four_point}

In case future PTA data will provide support for 
a  $2$-point ORF different
from the Hellings-Downs curve,  it will be crucial to design methods
for distinguishing among different explanations for  this phenomenon. We propose
a smoking-gun test for the non-Gaussian mechanism we outlined in the previous section, 
 extending to the PTA case the idea developed in  \cite{Seto:2009ju}
  in the context of ground-based GW detectors. 
 We consider  the following connected   4-point correlation function among PTA time-delay signals~\footnote{ It is also possible to consider 4-point functions between 4 pulsars, but  for simplicity we consider
only   two pulsars,  $\alpha$ and $\beta$.}
 $z_{\alpha, \beta}$ induced by GW
 \bea \label{defKab}
 {\cal K}_{\alpha\beta}(t_1,t_2)&\equiv&
 \langle z_{\alpha}(t_1)  z_{\alpha}(t_2) z_\beta(t_1) z_\beta(t_2) \rangle
 -\langle  
 z_{\alpha}(t_1)  z_{\alpha}(t_2) \rangle \langle z_\beta(t_1) z_\beta(t_2)\rangle
 \nonumber\\
 &-&
   \langle z_{\alpha}(t_1) z_\beta(t_1) \rangle  \langle z_{\alpha}(t_2) z_\beta(t_2) \rangle 
   - \langle z_{\alpha}(t_1) z_\beta(t_2) \rangle  \langle z_{\alpha}(t_2) z_\beta(t_1) \rangle 
  \label{defKabp}
  \,,
 \\
 &=&\frac{8 \pi}{3}\,\int d f\,e^{4 \pi i f\,   (t_1-t_2)}\,P(f)\,{ R}_{\alpha\beta}(f)
 \label{defKab}
 \,.
 \eea
The combination of the last three terms in eq \eqref{defKabp} is included in order to isolate
the connected
contribution
to the PTA $4$-point function,  depending   on the tensor  $4$-point function of eq \eqref{def4pfFa}.  
  In passing from eq \eqref{defKabp} to eq \eqref{defKab}, we make use of eq \eqref{statc1}.
 The quantity ${\cal K}_{\alpha\beta}(t_1,t_2)$ is stationary,
 since it depends on time differences only. We will learn that it is  non vanishing
 only in the presence of 4-point tensor non-Gaussianity, being it proportional to the quantities $\kappa_{1,2}$ entering in the
  4-point correlator of eqs   \eqref{def4pfFa}, \eqref{ans4ptFb}. We build  ${\cal K}_{\alpha\beta}(t_1,t_2)$ in terms
  of signals from two pulsars only, $\alpha$ and $\beta$ (instead
  of four distinct pulsars) for handling more easily the expressions involved,  and for being
  able to represent the corresponding ORF  in terms of a single angle $\zeta_{\alpha\beta}$. 

\smallskip

 The quantity ${ R}_{\alpha\beta}(f)$ is  the PTA  4-point ORF, and 
 can be expressed in terms of the quantity $x_{\alpha\beta}$ as defined in eq \eqref{defxab}, and of the 
 quantities $\kappa_{1,2}$ which characterize the polarization tensor $H_{\lambda_i}$
 given in eq \eqref{ans4ptFb}. 
 We find
 (see appendix \ref{app_A})
 \bea
 {R}_{\alpha\beta}&=&\frac{3}{40}\left(4 \kappa_1+\kappa_2 \right)-\frac{x_{\alpha\beta}}{40} \left(12 \kappa_1- 137 \kappa_2\right)
+\frac{x_{\alpha\beta}^2}{80} \left(4 \kappa_1-279 \kappa_2 \right)+\frac{3\,x_{\alpha\beta}\,\kappa_2}{2}
\left( 1+\frac32\,x_{\alpha\beta}
\right)\,\ln{x_{\alpha\beta}}\,,
\nonumber\\
 \label{rab4p}
 \eea
   In computing eq \eqref{rab4p}, we make use use only of the linear terms
   in the numerator of eq \eqref{mod2pt}, and neglect modulations induced by higher order, non-Gaussian ones (since  eq \eqref{rab4p} is already proportional to non-Gaussian contributions).

 
\begin{figure}
\centering
  \includegraphics[width = 0.6 \textwidth]{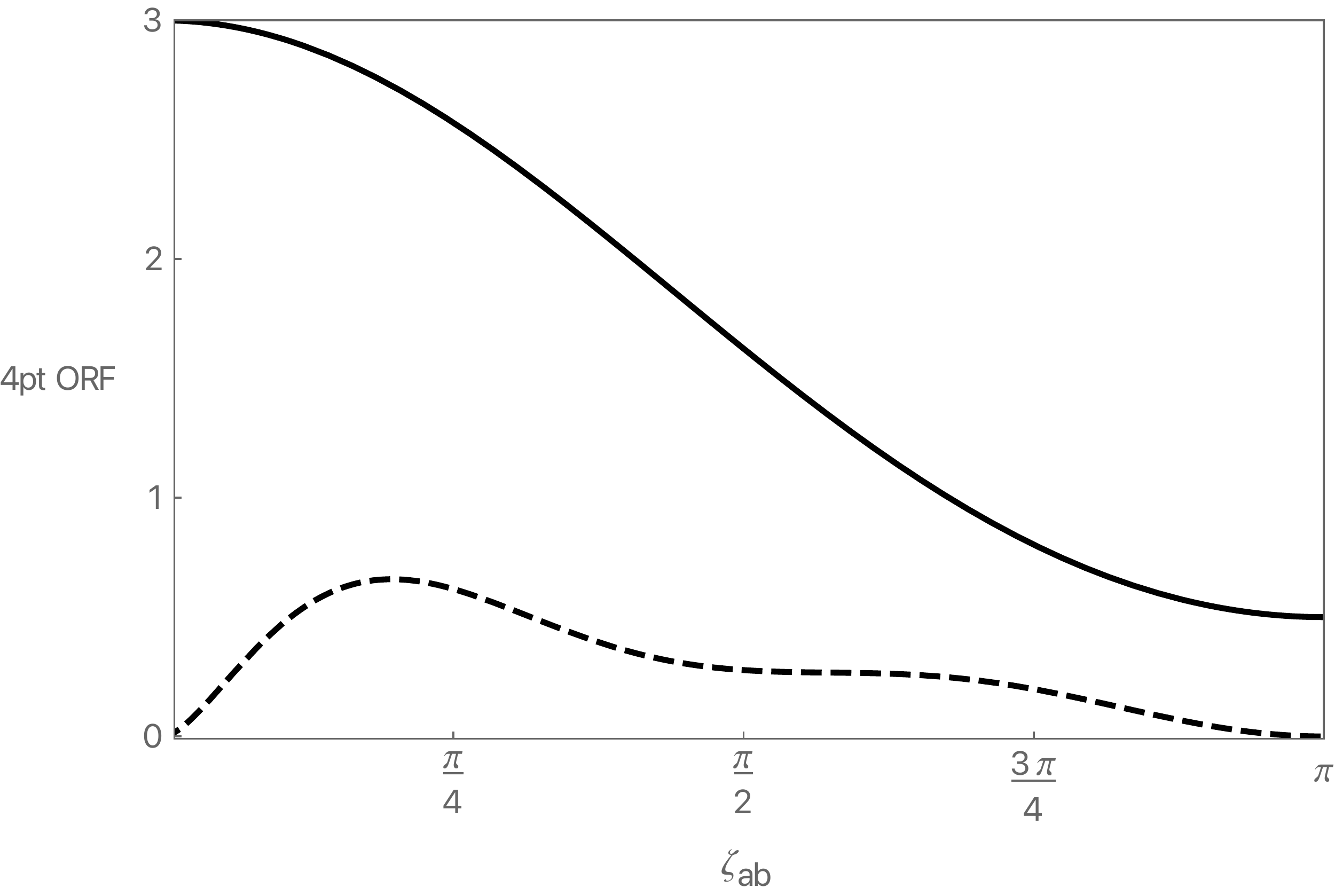}
 \caption{\small  The PTA 4-point overlap reduction function.  Solid
 line: $\kappa_1=10$, $\kappa_2=0$. Dashed
 line:  $\kappa_1=2$, $\kappa_2=-8$.}
 \label{fig:plot1aBc}
\end{figure}

  The corresponding ORF 
  is represented 
 in Fig \ref{fig:plot1aBc} for a representative choice of parameters. 
   Notice
  that,  depending on the relative size of $\kappa_{1,2}$, the amplitude of the ORF  $R_{\alpha\beta}$ can
  differ by around one order of magnitude for different choices of parameters. 
  In fact, 
  the angular dependence of the  4-point ORF is apparently  more sensitive to the helicity structure 
  of the  tensor $H_{\lambda_i}$ with respect to its 2-point counterpart
  of section \ref{sec_two_point}. 
    It would then be
  interesting to consider more general forms of $H_{\lambda_i}$
  than our Ansatz of eq \ref{ans4ptFb}, and study their consequences for the 4-point ORF.  
  

Let us quantify more the helicity dependence of the 4-point PTA overlap functions, with some simple statistical considerations. 
In appendix \ref{app_opt} we compute the optimal value of the signal-to-noise ratio (SNR) associated with a measurement
of the 4-point function of eq \eqref{defKab}, as summed over all the available
pulsar pairs. We find  the general expression
\be\label{opsnr1}
{\text{SNR}}^{\rm opt}\,=\,\sqrt{2 T}\,\left[ \frac{1}{S_n^4}\, 
\int d f \,|
 P(f)\,{ R}_{\rm tot}(f)|^2
\right]^{1/2}
\,,
\ee
where $T$ is the total duration of measurements, while 
 the pulsar noise $S_n$ 
is parametrized as  in eq \eqref{expcno}.  ${ R}_{\rm tot}$ is given in eq
 \eqref{defrtot} by the sum of the
4-point ORF $ R_{\alpha\beta}$ of eq \eqref{rab4p} when evaluated over all distinct pulsar pairs:
\be
{ R}_{\rm tot}(f)\,=\,\frac{8\pi}{3}\,\sum_{\alpha\beta} R_{\alpha\beta}
\ee
 The quantity ${ R}_{\rm tot}$ can
 enhance the SNR, if we have several pulsars to sum over: the result will also depend on the specific
 helicity structure of the quantity $H_{\lambda_i}$, in particular on the values of the parameters
 $\kappa_{1,2}$ when using Ansatz \eqref{ans4ptFb}.
   For the case of the  NANOGrav pulsars, using their angular position that can be extracted from Table 1 of  \cite{NANOGrav:2017wvv} we find the expression
\be
{ R}_{\rm tot}^{\rm NANOGrav}\,=\, 158.5\,\kappa_1+0.2\, \kappa_2
\ee
showing  that a NANOGrav  measurement of 4-point PTA correlations  would be more sensitive to the parameter  $\kappa_1$ with respect to $\kappa_2$. 

\section{Conclusions}
\label{sec_conclusions}

Several PTA collaborations are finding tantalising
 hints for  a SGWB
 signal in the nano-Hertz regime. So far, though, no 
 convincing evidence for   Hellings-Downs quadrupolar
 correlations has been found. While this issue can
 change at the light of more accurate, forthcoming
 data, it is meanwhile  important to explore scenarios able to 
 produce different types of  PTA angular correlations. We pointed
 out that a  stationary  non-Gaussian component to  the gravitational
 wave background can modulate the 2-point PTA overlap reduction
 function, modifying its angular dependence with additional parameters
 that might help in fitting data. We discussed possible
 sources for such non-Gaussian signal, as well as  additional 
 tests of this possibility.
 
 \smallskip
 
 Much questions are  left open in order to further elaborate  on this idea. It would be interesting
 to study in detail the shape and amplitude of GW
 non-Gaussianities
 in cosmological processes producing a large amplitude of SGWB after inflation ends, 
since these sources are able to produce the stationary non-Gaussianity we considered
in this work. Pursuing a complete analysis in realistic models would be helpful to determine
the frequency dependence and helicity structure of the $n$-point GW correlators that cause
the modulation of the PTA overlap functions. In order to perform proper fits with data, 
it would also be necessary to elaborate
 dedicated statistical analysis of signal+noise in the presence  of large non-Gaussianities in 
 the signal.
 
  Answering these questions will be interesting and compelling
 if future PTA data will   show evidence for a SGWB, but with  significant deviations from  Hellings-Downs angular 
 correlations.

\subsection*{Acknowledgments}

It is a pleasure to thank Debika Chowdhury, Emanuela Dimastrogiovanni, Matteo Fasiello, and Sachiko Kuroyanagi for discussions. GT is partially supported by the STFC grant ST/T000813/1.


\begin{appendix}

\section{Computation of the PTA time delay: including non-linearities}
\label{appTDPTA}

We extend the classic results of \cite{Detweiler:1979wn} for the computation of PTA time-delays in the presence
of GW, including effects of GW non-linearities.  
We express the space-time metric as (we set $c=1$)
\be
d s^2\,=\,-  d t^2+\left(\delta_{ij}+h_{ij} \right)\,d x^i d x^j \,,
\ee
with $h_{ij}$ the tensor fluctuation in transverse-traceless gauge.  We compute the time delay in the pulsar period 
due to the presence of a GW. We closely follow the textbook discussion of \cite{Maggiore:2018sht}, chapter 23, extending
it to the more general, non-linear  case we are interested in.

We  evaluate the  distance covered by photons travelling towards the earth, starting from pulsar $\alpha$
at spatial position $x_{\alpha}^i\,=\, x_\alpha\,\hat x^i_\alpha$,  with $\hat x^i_\alpha$ fixed
unit vector controlling the pulsar direction. Being the earth at $\vec x\,=\,0$, we
get the relation
\be \label{dxa1}
d x_\alpha^2\,=\,\frac{d t^2}{1+ h_{ij}\,\hat x^i_\alpha \hat x^j_\alpha }
\,,
\ee
controlling the infinitesimal geometrical distance covered by light during an interval  $dt$ in its way
from source to detection. For convenience, in what follows we assemble the 
combination $h_{ij}\,\hat x^i_\alpha \hat x^j_\alpha $ of eq \eqref{dxa1} into the quantity
\be
E_{\alpha}\,\equiv\,h_{ij}\,\hat x^i_\alpha \hat x^j_\alpha 
\ee
which   depends on time and space. We  assume that the GW,  controlled by $h_{ij}$, moves 
along a null-like geodesics,  and 
  has a characteristic frequency $\omega_{\rm GW}$ of the order of the inverse
of time it takes for light to arrive from source to detection. 

Photons emitted at time $t_{\rm em}$  are detected by an observer at time $t_{\rm obs}$ after
covering a comoving  distance
\bea
\label{defda1}
d_\alpha&=&\int_{t_{\rm em}}^{t_{\rm obs}}\,\frac{d t}{\left[
1+ E_\alpha
\right]^{1/2}}
\\
&=&t_{\rm obs}-t_{\rm em}+
\int_{t_{\rm em}}^{t_{\rm em}+d_\alpha+\delta t_o}\,d t'\,\left\{\frac{1}{\left[
1+ E_\alpha 
\right]^{1/2}}-1\right\}\left[t',\left( t_{\rm em}+ d_\alpha+\delta t_o-t'\right) \,\hat x_\alpha\right]
\nonumber
\,.
\\
\label{defda1a}
\eea
 Within the squared parentheses we have the coordinate dependence of the integrand function, which is  inside
the curly brackets. We use the fact  at time $t$  in the interval between $t_{\rm em}$ and $t_{\rm obs}$, photons lie at position $\vec x(t)\,=\,\left( t_{\rm obs}-t\right) \,\hat x_\alpha$. Moreover, since $h_{ij}$ is  small, in first approximation  we write $t_{\rm obs}\,=\,t_{\rm em}+ d_\alpha+\delta t_o$ in the integral, with $\delta t_o$ a small quantity depending on $E_\alpha $.  In the limit of vanishing
$h_{ij} $, we have $\delta t_o=0$. 

Suppose to consider a second train of photons emitted at a later  time $t_{\rm em}+T_\alpha$,
with $T_\alpha$ the pulsar period. We can then express the same quantity $d_\alpha$
of eq \eqref{defda1} as 
\bea
\label{defda2}
\hskip-0.2cm
d_\alpha&=&t'_{\rm obs}-t_{\rm em}-T_\alpha+
\int_{t_{\rm em}}^{t_{\rm em}+d_\alpha+\delta t'_o}\,d t'\,\left\{\frac{1}{\left[
1+ E_\alpha 
\right]^{1/2}}-1\right\}\left[t'+T_\alpha,\left( t_{\rm em}+ d_\alpha+\delta t'_o-t'\right) \,\hat x_\alpha\right]
\,,
\nonumber
\\
\eea
with $t'_{\rm obs}$ the new time of detection, and $\delta t_o'$ controls the difference, in the limit of small $h_{ij}$, between $t'_{\rm obs}$ and $t_{\rm em}+d_\alpha$. 

\smallskip

Taking the difference between \eqref{defda2} and \eqref{defda1a}, we find
\be
t'_{\rm obs}-t_{\rm obs}\,=\,T_\alpha+\Delta T_\alpha
\,,
\ee
with $\Delta T_\alpha$ given by the difference among the integrals appearing
in eqs \eqref{defda2} and \eqref{defda1a}. We know that the pulsar period is much smaller
than the time travel of light from source to detection.  This implies that the product $\omega_{\rm GW}\,T_\alpha$, which enters in the arguments of the time-dependent function
$E_{\alpha}$ in eq \eqref{defda2},  is  small.
 We can expand at first order in $T_\alpha$, finding the following expression for $\Delta T_\alpha$:
 \bea
\label{expDTA}
\Delta T_\alpha&=&\frac{T_\alpha}{2}
\,\int_{t_{\rm em}}^{t_{\rm em}+d_\alpha+\delta t_o}\,d t'\,\left[ \frac{\partial_{t'} E_{\alpha}(t', \vec x) }{\left( 1+ E_{\alpha}(t', \vec x) \right)^{3/2}}\right]_{\vec x\,=\,\vec x_0(t')}
\eea
with $\vec x_0(t)\,=\,t_{\rm em}+ d_\alpha+\delta t_o-t$.

   We assume that 
$E_{\alpha}$ can
 be modelled in terms of a monochromatic plane wave propagating in a null-like
 geodesics along the $\hat n$ direction:
 \be
 E_{\alpha}(t, \vec x)\,=\,E_{\alpha}\left( \omega_{\rm GW} (t-\hat n\,\vec x)\right)
 \,.
 \ee
We can  plug this expression in the integral of eq \eqref{expDTA}, and compute the time delay signal as
\bea
\label{finrtt1}
z_\alpha\,&\equiv&\frac{\Delta T_\alpha}{T_\alpha}
\,,
\nonumber
\\
&=&-\frac{1}{1+\hat x_\alpha \cdot \hat n}\left[\frac{1}{\sqrt{1+ E_{\alpha}(t, \vec x=0)}}
-\frac{1}{\sqrt{1+  E_{\alpha}(t-\tau_\alpha, \vec x=\vec x_\alpha)}}\right]
\,,
\eea
with $\tau_\alpha\,=\,t_{\rm em}+d_\alpha+\delta t_o$ the time travel from source to detection. This expression generalizes
the classic results of \cite{Detweiler:1979wn} including non-linearities in $E_{\alpha}$.  Expanding
up to third order in $E_{\alpha}$ we obtain  
eq \eqref{TaGEN1} in the main text.

\section{Computation of the PTA overlap reduction functions}
\label{app_A}

We denote with $\hat n$ the GW direction along the spatial
coordinates in a Cartesian system as $(\hat x,\,\hat y,\,\hat z)$.  We introduce two unit 
spatial vectors $\hat u$ and $\hat v$ orthogonal to $\hat n$:
\bea
\label{defhu}
\hat u&=&\frac{\hat n \,\times \,\hat z}{|\hat n \,\times \,\hat z|}
\,,
\\
\label{defhv}
\hat v&=&\frac{\hat n \,\times \,\hat u}{|\hat n \,\times \,\hat u|}
\,.
\eea
We can also express these quantities in spherical coordinates as
\bea
\hat n&=& \left( \sin{\theta} \cos\phi,\,\sin{\theta} \sin\phi,\,\cos{\theta}\right)
\,,
\\
\hat u&=& \left( \cos{\theta} \cos\phi,\,\cos{\theta} \sin\phi,\,-\sin{\theta}\right)
\,,
\\
\hat v&=&\left(\sin{\phi},\,-\cos{\phi},\,0 \right)
\,.
\eea

The symmetric $(+,\times)$ polarization tensors first introduced  in eq \eqref{sig_fourier} are defined as 
\bea
{\bf e}_{ij}^{(+)}&=&u_i u_j-v_i v_j
\,,
\\
{\bf e}_{ij}^{(\times)}&=&u_i v_j+v_i u_j
\,,
\eea
and satisfy the normalization condition
\be
{\bf e}_{ij}^{(\lambda_1)}\,{\bf e}_{ij}^{(\lambda_2)}\,=\,2 \delta^{\lambda_1 \lambda_2}
\,.
\ee

The two  vectors $\hat u $ and $\hat v$ introduced in eqs \eqref{defhu}, \eqref{defhv} are not the only unit vectors orthogonal to $\hat n$: more generally,  we can rotate $\hat u $ and $\hat v$ 
around $\hat n$ by an angle $\psi$:
\bea
\hat u'&=&\cos{\psi}\,\hat u+\sin{\psi}\,\hat v
\,,
\\
\hat v'&=&-\sin{\psi}\,\hat u+\cos{\psi}\,\hat v
\,.
\eea
Observables should not depend on $\psi$:  as in \cite{Seto:2009ju}, we   average over this
angle to determine $2$-point and  $4$-point overlap functions
used in the main text.

\smallskip
For computing the 2-point function, we introduce the quantities
\be
\bar E_\alpha^{(\lambda)}\,=\,{\bf e}^{(\lambda)}_{ij}\, \hat x_\alpha^i\,\hat x_\alpha^j
\,.
\ee
The pulsar positions are parametrized with $\hat x_\alpha\,=\,(0,0,1)$, $\hat x_\beta\,=\,(\cos \zeta_{\alpha\beta},0,\sin \zeta_{\alpha\beta})$. 
The correlators of eq \eqref{res2pt1}, once integrated over all GW directions, read
\bea
\langle z_\alpha(t, \vec x_\alpha) z_\beta (t, \vec x_\beta)\rangle
&=&
\frac{8\pi}{3}\int d f\,P(f)\,\Gamma_{\alpha \beta}(f)
\,.
\eea
 The 2-point overlap reduction function is  (as explained above, we average
over the angle $\psi$) 
\bea
\Gamma_{\alpha \beta}(f)&=&\frac{3}{32\pi^2}\,\sum_{\lambda_i}
\int_0^\pi d \psi\,\int_0^{2 \pi} d \phi\,\int_0^{\pi}\,\sin{\theta} d \theta\,\frac{1}{(1+\hat x_\alpha\cdot \hat n)} \,\frac{1}{(1+\hat x_\beta\cdot \hat n)} \times
\nonumber
\\
&&\times\left[\delta_{\lambda_1 \lambda_2}\,\bar E^{\lambda_1}_\alpha
\,\bar E^{\lambda_2}_\beta+H_{\lambda_1\lambda_2\lambda_3\lambda_4}
\left(\frac{9}{8} \bar E^{\lambda_1}_\alpha \bar E^{\lambda_2}_\alpha
\,\bar E^{\lambda_3}_\beta \,\bar E^{\lambda_4}_\beta +\frac54
 E^{\lambda_1}_\alpha  E^{\lambda_2}_\alpha
\,\bar E^{\lambda_3}_\alpha \,\bar E^{\lambda_4}_\beta
 +(\alpha\leftrightarrow\beta)\right) \right]
 \,,
 \nonumber\\
 &=&
 \left(1+\frac{29\, (\kappa_2+4 \kappa_1)}{140} \right)
 \nonumber\\
 &
 \times&
 \Big[
 \frac{140+116 \kappa_1+29 \kappa_2}{280}
-\frac{\left(140+988 \kappa_1+625 \kappa_2\right)}{560} x_{\alpha\beta}
+\frac{9}{280} \left(8\kappa_1+135 \kappa_2 \right)
x_{\alpha\beta}^2
\nonumber\\
&+&
\frac{\left(5492\kappa_1-2659\kappa_2\right)}{1120}
\,x_{\alpha\beta}^3+\frac{5(4\kappa_1+\kappa_2)}{4}\,x_{\alpha\beta}^4+\frac{3(4\kappa_1+\kappa_2)}{8}\,x_{\alpha\beta}^5
\nonumber\\
&+&
\frac32\,x_{\alpha\beta}\,\left( 1-9\kappa_1 \,x^2\right)\,\ln{x_{\alpha\beta}}
\Big]
\label{genorf2a}
\,,
\eea
where we used the tensor $H_{\lambda_1\dots\lambda_4}$ of eq \eqref{ans4ptFb}, and $x_{\alpha\beta}$ is defined in eq \eqref{defxab}. 
When setting $\kappa_1\,=\,\kappa_2\,=\,0$, we get the standard HD overlap reduction function. For $\kappa_2+4 \kappa_1\,=\,0$,
we get the function in eq \eqref{res2ptOV1} of the main draft. The angular integral can be done
straightforwardly, for example using the methods of \cite{Anholm:2008wy}. In this instance we used the residue theorem approach of \cite{Jenet:2014bea}. The overall coefficient of
eq \eqref{genorf2a} has been chosen such that the squared parenthesis approaches the value
$1/2$ at small values of $\zeta_{\alpha\beta}$, as the HD curve: we plot the  part inside the squared
parenthesis in Fig \ref{fig:plot1a}. The angular integration along $\psi$ plays an important role in the computation: we checked that only by using tensor 4-point functions one gets a non-vanishing result, while using 3-point functions one gets zero \cite{Seto:2009ju}.  

\bigskip
A very similar computation can be done for computing the 4-point overlap reduction
function discussed in section \ref{sec_four_point}. This quantity is given by (we sum over
repeated indexes)
\bea
\label{prorab1}
R_{\alpha\beta}(f)&=&\frac{3}{128\pi^2}\,
\int_0^\pi d \psi\,\int_0^{2 \pi} d \phi\,\int_0^{\pi}\,\sin{\theta} d \theta\,\frac{H_{\lambda_1\lambda_2\lambda_3\lambda_4}(f)
\,  \bar E^{\lambda_1}_\alpha \bar  E^{\lambda_2}_\alpha
\,\bar E^{\lambda_3}_\beta \,\bar E^{\lambda_4}_\beta }{(1+\hat x_\alpha\cdot \hat n)^2\, (1+\hat x_\beta\cdot \hat n)^2
}
\,,
\eea
and performing the angular integrations as above we get eq \eqref{rab4p} in the main text. 
   In computing $R_{\alpha\beta}(f)$ as in  eq \eqref{prorab1}, we make use use only of the linear terms
   in the numerator of eq \eqref{mod2pt}, and neglect modulations induced by higher order, non-Gaussian ones.

\section{Folded tensor  non-Gaussianity from causal  sources}
\label{app_model}

The aim of this appendix, following \cite{Green:2020whw},  is to  show through an explicit example  that tensor non-Gaussianities
from causal, classical sources can have enhanced support in a    folded  shape.  With `causal sources'  we refer to  contributions  from causal mechanisms active after inflation ends, for
example 
 associated 
with the decay of particles in their physical, initial state, due to non-linear 
interactions
that   respect locality and causality. As we will see, the corresponding $n$-point functions have poles
at  physical momenta, enhancing   folded non-Gaussian shapes.

 
 \smallskip
 
  The physically more interesting realisations of these  considerations  are
  associated with strong GW sources  which become active at 
   subhorizon scales after
 inflation ends -- i.e. during radiation and matter domination. 
  In such systems, the aforementioned locality and causality conditions are met.   Examples
 include phase transitions, or secondary sources of GWs associated
 with phenomena of PBH production. But, for ensuring that  our  arguments are as transparent as possible, 
   we focus in this appendix to  a  system in  pure de Sitter
 space, and select a specific local interaction for the tensor modes: essentially, we apply the arguments 
 of \cite{Green:2020whw} to the tensor case.  
 
 \smallskip
 
 We start  with an explicit computation of connected  $4$-point function of 
 tensor modes using as  \cite{Green:2020whw} the method of Green functions, to then discuss its physical consequences. 
 Besides the usual free quadratic action for spin-2 tensor fluctuations in de Sitter space, we consider
  a representative local quartic interaction described by the Hamiltonian density
 \be \label{quarint1}
  {\cal H}_{\rm int}\,=\,-\frac{q_0}{4!}\,\dot h_{ij}^4\,,
 \ee
 with $q_0$ a constant, and for simplicity we neglect cubic interactions, since we will focus on 4-point
 correlators. The interaction
 \eqref{quarint1} allows for
 tensor fluctuations in their physical state to decay (or annihilate) on-shell through a non-linear $1\leftrightarrow3$ process,
 leading -- as we will see -- to a characteristic pole structure in the $4$-point correlation functions.
  
 
 \smallskip
 
  To make more direct connection with standard computations of $4$-point
 correlation functions in field theory, in this appendix we Fourier expand
 the spin-2 field implementing  a slightly different notation with respect to the main text.
 We work in conformal time, $\partial_t\,=\,a^{-1}(\tau)\,\partial_\tau$, and write 
  

 \be
 h_{ij}(\tau, \vec x)\,=\,\sum_\lambda\int \frac{d^3 k}{(2 \pi)^3}\,e^{i \vec k \vec x}
 \,{\bf e}^{(\lambda)}_{ij}(\hat k)
\,\tilde h_\lambda(\tau, \vec k)
\,,
 \ee 
 where the Fourier mode is decomposed in terms of classical stochastic quantities  as
 \be
\tilde h_\lambda(\tau, \vec k)\,=\,
 \hat a_{\lambda}^\dagger( \vec k)\,\bar h_{
 k}(\tau)+\hat a_{\lambda}( -\vec k)\,\bar h_{k}^*(\tau)
 \,,
 \ee
 and we denote $\hat k\,\equiv\,\vec k/| \vec k|$, and $k\equiv |\vec k|$.
The stochastic parameters $\hat a_{\lambda}^\dagger$, $\hat a_{\lambda}$ are classical, commuting 
quantities satisfying the  statistical conditions
\be
\langle \hat a_{\lambda_1}^\dagger({\vec k}_1) \, \hat a_{\lambda_2}({\vec k}_2) \rangle\,=\,\frac12
\delta (\vec k_1-\vec k_2)\,\delta_{\lambda_1 \lambda_2}\,=\,\langle \hat a_{\lambda_1}({\vec k}_2) \, \hat a_{\lambda_2}^\dagger({\vec k}_1) \rangle
\,,
\ee
as ensemble averages -- see  \cite{Green:2020whw}.
 The solution for the linearized mode function $\bar h_{
 \lambda, k}(\tau)$ in de Sitter space is
\be
\bar h_{  k}(\tau)\,=\,\frac{\Delta_h}{k^{3/2}}\,e^{i k \tau}\, (1-i k \tau)
\,,
\ee 
with $\Delta_h$ a constant quantity controlling  the spin-2 normalization.  These results imply that,
working at the linearized level, the equal-time spin-2 correlation functions satisfy 
\be
\langle  \tilde h^{(1)}_{\lambda_1}(\tau, \vec k_1)\, \tilde h^{(1)}_{\lambda_2}(\tau, \vec k_2)
\rangle\,=\,\delta (\vec k_1+\vec k_2)\,\delta_{\lambda_1 \lambda_2}\,\frac{\Delta_h^2}{k_1^3}\,\left(1+k_1^2 \tau^2\right)
\,.
\ee
We now proceed including the effects of interactions. 
We start from the evolution equations
for tensor modes including the quartic interaction \eqref{quarint1} (primes
denote derivatives along time):
\be \label{eveq1rs}
  h''_{ij}+2 {\cal H}\,h'_{ij}-{\nabla^2} h_{ij}\,=\,\frac{q_0}{12\,a^2} \partial_\tau \left(  \partial_\tau h_{ij} \right)^3
  \,.
  \ee
 Using the properties of the polarization tensors  we  can rewrite eq \eqref{eveq1rs}  in Fourier space as 
  \bea
&&\tilde h''_{\lambda}(\tau,\vec k)+2 {\cal H}\,\tilde h'_{\lambda}(\tau,\vec k)+k^2\,\tilde h_{\lambda}(\tau,\vec k)
\nonumber
\\
&&\,=\,\frac{q_0}{24\,a^2}\,\partial_\tau \left[ \int \frac{d^3  q_1}{(2 \pi)^3} \frac{d^3  q_2}{(2 \pi)^3} 
 \, H_\lambda^{ \lambda_a \lambda_b \lambda_c}   \tilde h'_{\lambda_a}(\tau, \vec  q_1)  \tilde h'_{\lambda_b}(\tau, \vec  q_2)
 \,  \tilde h'_{\lambda_c}(\tau, \vec k-\vec q_1-\vec q_2)
 \right]
  \,,
  \eea
  with
  \be
 H^{\lambda \lambda_a \lambda_b \lambda_c}\,=\,{\bf e}_{ij}^{(\lambda)}
(\hat k) {\bf e}_{im}^{(\lambda_a)}(\hat q_1)  {\bf e}_{mn}^{(\lambda_b)}(\hat q_2)
{\bf e}_{nj}^{(\lambda_c)}(\hat   k-\hat q_1-\hat q_2)
  \,.
  \ee

  Following \cite{Green:2020whw} 
we can use the Green function method for studying the effects of classical non-linearities, and how
they source connected   $n$-point correlation functions.  
 The Green function $G_k(\tau, \tau')$ relative to  the spin-2 evolution equation  \eqref{eveq1rs} in pure de Sitter 
 can be expressed as 
 \be
G_k(\tau, \tau')\,=\,\frac{2\,\Delta_h^2}{k^3}\,\left\{  \sin{\left[k (\tau-\tau')\right]}\,\left(1+k^2 \tau \tau'\right)
-k\,\left( \tau-\tau'\right) \,\cos{\left[k (\tau-\tau')\right]}
\right\}
\label{resGF1}
 \,.
 \ee
 We can decompose the tensor fluctuation in a linear and cubic term
 in 
  momentum space ($\tilde h_{\lambda}(\tau, \vec k)\,=\,\tilde h^{(1)}_{\lambda}(\tau, \vec k)+\tilde h^{(3)}_{\lambda}(\tau, \vec k)$). 
  By using the
 %
 %
   Green function of eq \eqref{resGF1},
 the formal expression for the spin-2 solution  at third order is (the sum over repeated indexes is understood)
 \bea
 \hskip-0.3cm
\tilde h_{\lambda}^{(3)}(\tau, \vec k)&=&\frac{q_0}{24}\,
\int d \tau' \frac{d^3  q_1}{(2 \pi)^3} \frac{d^3  q_2}{(2 \pi)^3} 
H_\lambda^{ \lambda_a \lambda_b \lambda_c}
\partial_{\tau'}G_k(\tau, \tau') \,\partial_{\tau'} \tilde h^{(1)}_{\lambda_a}(\tau',  \vec q_1)\,
  \partial_{\tau'} \tilde h^{(1)}_{\lambda_b}(\tau', \vec  q_2)
 \, \partial_{\tau'}  \tilde h^{(1)}_{\lambda_c}(\tau', \vec k-\vec q_1-\vec q_2)
\,.
\nonumber\\
\eea
This expression can be used to compute the connected part of the equal-time  4-point correlation
function of $\tilde  h_{\lambda}(\tau, \vec k) $ at leading order in $q_0$. We find (understanding the equal time dependence)
\bea
&&
\langle 
\tilde
h_{\lambda_1}(\vec k_1)
\tilde h_{\lambda_2}( \vec k_2)
\tilde h_{\lambda_3}( \vec k_3)
\tilde h_{\lambda_4}( \vec k_4)
\rangle\,=\,
\nonumber
\\
&&
\hskip-1.1cm \frac{q_0}{4}\,\int\,d \tau'
\frac{d^3  q_1}{(2 \pi)^3} \frac{d^3  q_2}{(2 \pi)^3} 
 \,H^{\lambda_4}_{ \lambda_a \lambda_b \lambda_c}\,(\partial_{\tau'}G_{k_4})\,
\langle\tilde
h_{\lambda_1}( \vec k_1)
 \tilde h_{\lambda_a}( \vec q_1)
\rangle \langle \tilde
h_{\lambda_2}( \vec k_2) \tilde
h_{\lambda_b)}( \vec q_2) \rangle  \langle  \tilde h_{\lambda_3}( \vec k_3)\,\tilde h_{\lambda_c}(\vec k_4-\vec q_1-\vec q_2)
\rangle
\nonumber
\\
&&
 \hskip-0.9cm
 =\frac{9 \,q_0}{4}\, \delta(\vec k_1+\vec k_2+ \vec k_3 +\vec k_4)
\,\frac{\Delta_h^8}{k_1 k_2 k_3 k_4} \,H_{\lambda_4 \lambda_1 \lambda_2 \lambda_3}\,
\nonumber
\\
&&
 \hskip-0.9cm
\times
\Big[
\frac{1}{\left( k_1+k_2+k_3+k_4\right)^5}+\frac{1}{\left( k_4-k_1-k_2-k_3\right)^5}
+\frac{1}{\left( k_1+k_2-k_3+k_4\right)^5}
+\frac{1}{\left( k_2+k_3+k_4-k_1\right)^5}
\nonumber\\
 &&
  \hskip-0.7cm
 +\frac{1}{\left( k_1+k_3+k_4-k_2\right)^5}
+\frac{1}{\left( k_3+k_4-k_1-k_2\right)^5}+\frac{1}{\left( k_1+k_3-k_4-k_2\right)^5}
+\frac{1}{\left( k_1+k_4-k_3-k_2\right)^5}
 \Big]
  +{\rm perms}\,.
  \nonumber
 \\
\label{bigr4pt}
\eea
where
\be
H_{\lambda_4 \lambda_1 \lambda_2 \lambda_3}\,=\,{\bf e}_{ij}^{(\lambda_4)}
(\hat k_4) {\bf e}_{im}^{(\lambda_1)}(-\hat k_1)  {\bf e}_{mn}^{(\lambda_2)}(-\hat k_2)
{\bf e}_{nj}^{(\lambda_3)}(-\hat   k_1-\hat k_2-\hat k_4)
\,.
\ee

The tensor 4-point function \eqref{bigr4pt} contains poles at physical momenta, which enhance
a folded shape of tensor non-Gaussianity,  corresponding to a quadrangle with superimposed sides 
in Fourier space. This example shows that classical correlators from causal sources
provide the shape of non-Gaussian signals we are after, and which can   source the effects
discussed in section \ref{sec_two_point}. Notice that the divergences at the poles can be smoothed
by effects as classical dissipation \cite{Green:2020whw}: nevertheless, the corresponding correlation
functions  have most of their support in folded shapes.

\section{The optimal signal-to-noise ratio for the PTA  $4$-point function}
\label{app_opt}

We   determine the optimal signal-to-noise
ratio (SNR) for estimating the stationary $4$-point function considered
in section \ref{sec_four_point}.
 We generalize the arguments of  \cite{Powell:2019kid}, which uses  methods
 developed in \cite{Anholm:2008wy,Thrane:2013oya} and reviewed in \cite{Maggiore:2007ulw}.  
 We assume that the time-delay signal $s_\alpha$ measured with pulsar
 experiments can be  separated in a `true' GW signal $z_\alpha$ 
 (as given in eq \eqref{defKab}) and
 uncorrelated noise $n_\alpha$:
 \be
 s_\alpha\,=\,z_\alpha+n_\alpha
 \,.
 \ee 
 We then integrate 
 the stationary $4$-point correlator among signals from
 two pulsars $\alpha$ and $\beta$ over the temporal
 duration $T$ of the experiment, and we  
  define the quantity ${\cal Y}_{\alpha \beta}$:
\be
\label{defYab1}
{\cal Y}_{\alpha \beta}\,=\,\int_{-T/2}^{T/2} d t_1
\,
d t_2
\,{\cal K}_{\alpha\beta}(t_1, t_2)\,{\cal F}(t_2-t_1)
\,,
\ee
where ${\cal K}_{\alpha\beta}(t_1, t_2)$, as in eq \eqref{defKab},  is  a product of four signals evaluated
at two different times, as measured at the earth:
\bea
{\cal K}_{\alpha\beta}(t_1, t_2)&=& s_{\alpha}(t_1)  s_{\alpha}(t_2) s_\beta(t_1) s_\beta(t_2) 
\,.
\eea
 The function ${\cal F}$ in eq \eqref{defYab1} is  a yet-to-be-determined
 filter function which decays rapidly with increasing the size of
 its argument $|t_i-t_j|$.

  In defining  the SNR$=S/N$, the quantity $S$ corresponds to the
  connected part of the ensemble average value of ${\cal Y}_{\alpha \beta}$ in the presence of the GW signal (see
  eq \eqref{defKab}); the noise $N$  is the root mean square value of ${\cal Y}_{\alpha \beta}$ when the signal is absent. We   determine the  filter function ${\cal F}$  that maximises the corresponding SNR. 
 We Fourier transform \eqref{defYab1}, finding
 \bea
 {\cal Y}_{\alpha\beta}&=&
 \int_{-\infty}^{\infty}
 d f_A\,d f_B\,d f_C\,
  \,\delta_T(f_A+f_C) \,\delta_T(f_C-f_B)
  \,\tilde{\cal F}(f_C)
 \,\tilde {\cal K}_{\alpha\beta}(f_A, f_B)
 \,,
 \label{defYabAA}
 \eea
and 
we introduce   
$\delta_T(f)\equiv\int_{-T}^T\,\exp{\left[2 \pi i f t
\right]}\,d t$, a function  with the property $\delta_T(0)\,=\,T$. 
Eq \eqref{defYabAA} is the starting point for our computations of $S$ and $N$. 

\smallskip
For the signal $S$ we  use the stationary property \eqref{defKab} characterizing the connected GW
$4$-point functions, which implies\footnote{The factors
of $1/2$ in the arguments of the functions are due to the $e^{4\pi i f(t_1-t_2)}$ factor 
in  eq \eqref{defKab}.}
\be
\tilde {\cal K}_{\alpha\beta}(f_A, f_B)\,=\,\frac{8 \pi}{3}\,\delta(f_A+f_B)\,
  P(f_A/2)\,{ R}_{\alpha\beta}(f_A/2)
  \,,
\ee
with $R_{\alpha\beta}$ given in eq \eqref{rab4p}. 
Plugging this expression in eq \eqref{defYabAA}, we find that the `signal' contribution
is 
\bea
 S&=&\frac{8 \pi}{3}
 \int_{-\infty}^{\infty}
 d f_A\,d f_B\,d f_C\,
  \,\delta_T(f_A+f_C) \,\delta_T(f_C-f_B)
  \,\tilde{\cal F}(f_C)\,
\delta(f_A+f_B)\,
 P(f_A/2)\,{ R}_{\alpha\beta}(f_A/2)
  \nonumber
  \\
  &=&\frac{8 \pi\,T}{3}
\,\,
  \int_{-\infty}^{\infty}
 d f
  \,\tilde{\cal F}(f)\,
 P(f/2)\,{ R}_{\alpha\beta}(f/2)
 \,.
  \label{defYabAA1}
 \eea
 
 \smallskip
 We  now consider the noise part. 
 We assume the noise has a Gaussian distribution, with 2-point
 correlation function 
 \be
\langle n_\alpha(t_1)\,n_{\beta}(t_2) \rangle\,=\,S_n\,\delta(t_1-t_2)\,\delta_{\alpha\beta}
\,.
\ee
For simplicity we assume a common $S_n$ for all pulsars, that as 
\cite{Thrane:2013oya} we  parametrize as
\be
\label{expcno}
S_n\,=\,2 
\,
\Delta t
\,\sigma^2
\,,
\ee
with $1/\Delta t$ the typical measurement cadence, and $\sigma^2$ the 
{\it rms} of the noise timing. 
The noise results
\bea
N^2&=&
\langle {\cal Y}_{\alpha\beta} {\cal Y}_{\alpha\beta} \rangle \,=\,
T\,S_n^4\,\int d f \,|\tilde {\cal F}(f)|^2\,.
\label{fenoi}
\eea
We can then build the total SNR assembling the results of eq \eqref{defYabAA1}
 and eq
\eqref{fenoi}, summing over all the pulsar
pairs, and denoting for brevity
 \be
 \label{defrtot}
{ R}_{\rm tot}(f)\,\equiv\,\frac{8 \pi}{3}\,\sum_{\alpha\beta} { R}_{\alpha\beta}(f)
\,.
\ee
We find
\be
\label{finsnr}
{\rm SNR}\,=\,\sqrt{T}\,\frac{ \int d f 
  \,\tilde{\cal F}(f)\,
 P(f/2)\,{ R}_{\rm tot}(f/2)}{S_n^2\,\left[\int d f \,|\tilde {\cal F}(f)|^2\right]^{1/2}}
 \,.
\ee
 It is easy to determine the filter function $\tilde {\cal F}$ that maximizes the previous expression. 
We introduce a positive definite scalar product between two arbitrary quantities $A_i(f)$
 \be
[ A_1(f), A_2(f)]\,\equiv\,\int d f\,A_1(f)\,A_2^\star(f)\,S_n^4
\,.
\ee
Then the SNR of eq \eqref{finsnr} can be schematically expressed as
\be
{\text{SNR}}\,=\,\sqrt{T}\,\frac{[{\tilde{\cal F}}(f) , 
\, P(f/2)\,{ R}_{\rm tot}(f/2)/S_n^4] }{[{\tilde{\cal F}}(f)
 ,{\tilde{\cal F}}(f)]^{1/2}}
 \,,
\ee
and it is maximised by choosing an optimal filter function such that
\be
{\tilde{\cal F}}(f) \,=\,
 P(f/2)\,{ R}_{\rm tot}(f/2)/S_n^4
 \,.
\ee
Plugging this result  in eq \eqref{finsnr}, we find that the optimal SNR results
\be\label{opsnr}
{\text{SNR}}^{\rm opt}\,=\,\sqrt{2 T}\,\left[ \frac{1}{S_n^4}\, 
\int d f \,|
 P(f)\,{ R}_{\rm tot}(f)|^2
\right]^{1/2}
\,.
\ee
The result depends both on the values of the 4-pt correlation of the GW signal, and
on the location of pulsars entering in the quantity ${ R}_{\rm tot}$.

\end{appendix}
\providecommand{\href}[2]{#2}\begingroup\raggedright\endgroup


\begin{thebibliography}{10}

\bibitem{Sazhin:OO}
M.~Sazhin, ``{Opportunities for detecting ultralong gravitational waves},''
  \href{http://dx.doi.org/1978SvA....22...36S}{{\em Soviet Astronomy}
  {\bfseries 22} (1978) 36--38}.

\bibitem{Detweiler:1979wn}
S.~L. Detweiler, ``{Pulsar timing measurements and the search for gravitational
  waves},'' \href{http://dx.doi.org/10.1086/157593}{{\em Astrophys. J.}
  {\bfseries 234} (1979) 1100--1104}.

\bibitem{Mashhoon:1979wk}
B.~Mashhoon, ``{ON THE DETECTION OF GRAVITATIONAL RADIATION BY THE DOPPLER
  TRACKING OF SPACECRAFT},'' \href{http://dx.doi.org/10.1086/156810}{{\em
  Astrophys. J.} {\bfseries 227} (1979) 1019--1036}.

\bibitem{Bertotti:1980pg}
B.~Bertotti and B.~J. Carr, ``{THE PROSPECTS OF DETECTING GRAVITATIONAL
  BACKGROUND RADIATION BY DOPPLER TRACKING INTERPLANETARY SPACECRAFT},''
  \href{http://dx.doi.org/10.1086/157826}{{\em Astrophys. J.} {\bfseries 236}
  (1980) 1000--1011}.

\bibitem{Lommen:2015gbz}
A.~N. Lommen, ``{Pulsar timing arrays: the promise of gravitational wave
  detection},'' \href{http://dx.doi.org/10.1088/0034-4885/78/12/124901}{{\em
  Rept. Prog. Phys.} {\bfseries 78} no.~12, (2015) 124901}.

\bibitem{NANOGrav:2020bcs}
{\bfseries NANOGrav} Collaboration, Z.~Arzoumanian {\em et~al.}, ``{The
  NANOGrav 12.5 yr Data Set: Search for an Isotropic Stochastic
  Gravitational-wave Background},''
  \href{http://dx.doi.org/10.3847/2041-8213/abd401}{{\em Astrophys. J. Lett.}
  {\bfseries 905} no.~2, (2020) L34},
  \href{http://arxiv.org/abs/2009.04496}{{\ttfamily arXiv:2009.04496
  [astro-ph.HE]}}.

\bibitem{Goncharov:2021oub}
B.~Goncharov {\em et~al.}, ``{On the Evidence for a Common-spectrum Process in
  the Search for the Nanohertz Gravitational-wave Background with the Parkes
  Pulsar Timing Array},''
  \href{http://dx.doi.org/10.3847/2041-8213/ac17f4}{{\em Astrophys. J. Lett.}
  {\bfseries 917} no.~2, (2021) L19},
  \href{http://arxiv.org/abs/2107.12112}{{\ttfamily arXiv:2107.12112
  [astro-ph.HE]}}.

\bibitem{Chen:2021rqp}
S.~Chen {\em et~al.}, ``{Common-red-signal analysis with 24-yr high-precision
  timing of the European Pulsar Timing Array: inferences in the stochastic
  gravitational-wave background search},''
  \href{http://dx.doi.org/10.1093/mnras/stab2833}{{\em Mon. Not. Roy. Astron.
  Soc.} {\bfseries 508} no.~4, (2021) 4970--4993},
  \href{http://arxiv.org/abs/2110.13184}{{\ttfamily arXiv:2110.13184
  [astro-ph.HE]}}.

\bibitem{Antoniadis:2022pcn}
J.~Antoniadis {\em et~al.}, ``{The International Pulsar Timing Array second
  data release: Search for an isotropic Gravitational Wave Background},''
  \href{http://dx.doi.org/10.1093/mnras/stab3418}{{\em Mon. Not. Roy. Astron.
  Soc.} {\bfseries 510} no.~4, (2022) },
  \href{http://arxiv.org/abs/2201.03980}{{\ttfamily arXiv:2201.03980
  [astro-ph.HE]}}.

\bibitem{Haehnelt:1994wt}
M.~G. Haehnelt, ``{Low frequency gravitational waves from supermassive black
  holes},'' \href{http://dx.doi.org/10.1093/mnras/269.1.199}{{\em Mon. Not.
  Roy. Astron. Soc.} {\bfseries 269} (1994) 199},
  \href{http://arxiv.org/abs/astro-ph/9405032}{{\ttfamily
  arXiv:astro-ph/9405032}}.

\bibitem{Sesana:2004sp}
A.~Sesana, F.~Haardt, P.~Madau, and M.~Volonteri, ``{Low - frequency
  gravitational radiation from coalescing massive black hole binaries in
  hierarchical cosmologies},'' \href{http://dx.doi.org/10.1086/422185}{{\em
  Astrophys. J.} {\bfseries 611} (2004) 623--632},
  \href{http://arxiv.org/abs/astro-ph/0401543}{{\ttfamily
  arXiv:astro-ph/0401543}}.

\bibitem{Sesana:2008mz}
A.~Sesana, A.~Vecchio, and C.~N. Colacino, ``{The stochastic gravitational-wave
  background from massive black hole binary systems: implications for
  observations with Pulsar Timing Arrays},''
  \href{http://dx.doi.org/10.1111/j.1365-2966.2008.13682.x}{{\em Mon. Not. Roy.
  Astron. Soc.} {\bfseries 390} (2008) 192},
  \href{http://arxiv.org/abs/0804.4476}{{\ttfamily arXiv:0804.4476
  [astro-ph]}}.

\bibitem{Vaskonen:2020lbd}
V.~Vaskonen and H.~Veerm\"ae, ``{Did NANOGrav see a signal from primordial
  black hole formation?},''
  \href{http://dx.doi.org/10.1103/PhysRevLett.126.051303}{{\em Phys. Rev.
  Lett.} {\bfseries 126} no.~5, (2021) 051303},
  \href{http://arxiv.org/abs/2009.07832}{{\ttfamily arXiv:2009.07832
  [astro-ph.CO]}}.

\bibitem{DeLuca:2020agl}
V.~De~Luca, G.~Franciolini, and A.~Riotto, ``{NANOGrav Data Hints at Primordial
  Black Holes as Dark Matter},''
  \href{http://dx.doi.org/10.1103/PhysRevLett.126.041303}{{\em Phys. Rev.
  Lett.} {\bfseries 126} no.~4, (2021) 041303},
  \href{http://arxiv.org/abs/2009.08268}{{\ttfamily arXiv:2009.08268
  [astro-ph.CO]}}.

\bibitem{Kohri:2020qqd}
K.~Kohri and T.~Terada, ``{Solar-Mass Primordial Black Holes Explain NANOGrav
  Hint of Gravitational Waves},''
  \href{http://dx.doi.org/10.1016/j.physletb.2020.136040}{{\em Phys. Lett. B}
  {\bfseries 813} (2021) 136040},
  \href{http://arxiv.org/abs/2009.11853}{{\ttfamily arXiv:2009.11853
  [astro-ph.CO]}}.

\bibitem{Ellis:2020ena}
J.~Ellis and M.~Lewicki, ``{Cosmic String Interpretation of NANOGrav Pulsar
  Timing Data},'' \href{http://dx.doi.org/10.1103/PhysRevLett.126.041304}{{\em
  Phys. Rev. Lett.} {\bfseries 126} no.~4, (2021) 041304},
  \href{http://arxiv.org/abs/2009.06555}{{\ttfamily arXiv:2009.06555
  [astro-ph.CO]}}.

\bibitem{Buchmuller:2020lbh}
W.~Buchmuller, V.~Domcke, and K.~Schmitz, ``{From NANOGrav to LIGO with
  metastable cosmic strings},''
  \href{http://dx.doi.org/10.1016/j.physletb.2020.135914}{{\em Phys. Lett. B}
  {\bfseries 811} (2020) 135914},
  \href{http://arxiv.org/abs/2009.10649}{{\ttfamily arXiv:2009.10649
  [astro-ph.CO]}}.

\bibitem{Blasi:2020mfx}
S.~Blasi, V.~Brdar, and K.~Schmitz, ``{Has NANOGrav found first evidence for
  cosmic strings?},''
  \href{http://dx.doi.org/10.1103/PhysRevLett.126.041305}{{\em Phys. Rev.
  Lett.} {\bfseries 126} no.~4, (2021) 041305},
  \href{http://arxiv.org/abs/2009.06607}{{\ttfamily arXiv:2009.06607
  [astro-ph.CO]}}.

\bibitem{Blanco-Pillado:2021ygr}
J.~J. Blanco-Pillado, K.~D. Olum, and J.~M. Wachter, ``{Comparison of cosmic
  string and superstring models to NANOGrav 12.5-year results},''
  \href{http://dx.doi.org/10.1103/PhysRevD.103.103512}{{\em Phys. Rev. D}
  {\bfseries 103} no.~10, (2021) 103512},
  \href{http://arxiv.org/abs/2102.08194}{{\ttfamily arXiv:2102.08194
  [astro-ph.CO]}}.

\bibitem{Nakai:2020oit}
Y.~Nakai, M.~Suzuki, F.~Takahashi, and M.~Yamada, ``{Gravitational Waves and
  Dark Radiation from Dark Phase Transition: Connecting NANOGrav Pulsar Timing
  Data and Hubble Tension},''
  \href{http://dx.doi.org/10.1016/j.physletb.2021.136238}{{\em Phys. Lett. B}
  {\bfseries 816} (2021) 136238},
  \href{http://arxiv.org/abs/2009.09754}{{\ttfamily arXiv:2009.09754
  [astro-ph.CO]}}.

\bibitem{Ratzinger:2020koh}
W.~Ratzinger and P.~Schwaller, ``{Whispers from the dark side: Confronting
  light new physics with NANOGrav data},''
  \href{http://dx.doi.org/10.21468/SciPostPhys.10.2.047}{{\em SciPost Phys.}
  {\bfseries 10} no.~2, (2021) 047},
  \href{http://arxiv.org/abs/2009.11875}{{\ttfamily arXiv:2009.11875
  [astro-ph.CO]}}.

\bibitem{Addazi:2020zcj}
A.~Addazi, Y.-F. Cai, Q.~Gan, A.~Marciano, and K.~Zeng, ``{NANOGrav results and
  dark first order phase transitions},''
  \href{http://dx.doi.org/10.1007/s11433-021-1724-6}{{\em Sci. China Phys.
  Mech. Astron.} {\bfseries 64} no.~9, (2021) 290411},
  \href{http://arxiv.org/abs/2009.10327}{{\ttfamily arXiv:2009.10327
  [hep-ph]}}.

\bibitem{NANOGrav:2021flc}
{\bfseries NANOGrav} Collaboration, Z.~Arzoumanian {\em et~al.}, ``{Searching
  for Gravitational Waves from Cosmological Phase Transitions with the NANOGrav
  12.5-Year Dataset},''
  \href{http://dx.doi.org/10.1103/PhysRevLett.127.251302}{{\em Phys. Rev.
  Lett.} {\bfseries 127} no.~25, (2021) 251302},
  \href{http://arxiv.org/abs/2104.13930}{{\ttfamily arXiv:2104.13930
  [astro-ph.CO]}}.

\bibitem{Brandenburg:2021tmp}
A.~Brandenburg, E.~Clarke, Y.~He, and T.~Kahniashvili, ``{Can we observe the
  QCD phase transition-generated gravitational waves through pulsar timing
  arrays?},'' \href{http://dx.doi.org/10.1103/PhysRevD.104.043513}{{\em Phys.
  Rev. D} {\bfseries 104} no.~4, (2021) 043513},
  \href{http://arxiv.org/abs/2102.12428}{{\ttfamily arXiv:2102.12428
  [astro-ph.CO]}}.

\bibitem{Neronov:2020qrl}
A.~Neronov, A.~Roper~Pol, C.~Caprini, and D.~Semikoz, ``{NANOGrav signal from
  magnetohydrodynamic turbulence at the QCD phase transition in the early
  Universe},'' \href{http://dx.doi.org/10.1103/PhysRevD.103.L041302}{{\em Phys.
  Rev. D} {\bfseries 103} no.~4, (2021) 041302},
  \href{http://arxiv.org/abs/2009.14174}{{\ttfamily arXiv:2009.14174
  [astro-ph.CO]}}.

\bibitem{RoperPol:2022iel}
A.~Roper~Pol, C.~Caprini, A.~Neronov, and D.~Semikoz, ``{The gravitational wave
  signal from primordial magnetic fields in the Pulsar Timing Array frequency
  band},'' \href{http://arxiv.org/abs/2201.05630}{{\ttfamily arXiv:2201.05630
  [astro-ph.CO]}}.

\bibitem{Hellings:1983fr}
R.~w. Hellings and G.~s. Downs, ``{UPPER LIMITS ON THE ISOTROPIC GRAVITATIONAL
  RADIATION BACKGROUND FROM PULSAR TIMING ANALYSIS},''
  \href{http://dx.doi.org/10.1086/183954}{{\em Astrophys. J. Lett.} {\bfseries
  265} (1983) L39--L42}.

\bibitem{Chen:2021wdo}
Z.-C. Chen, C.~Yuan, and Q.-G. Huang, ``{Non-tensorial gravitational wave
  background in NANOGrav 12.5-year data set},''
  \href{http://dx.doi.org/10.1007/s11433-021-1797-y}{{\em Sci. China Phys.
  Mech. Astron.} {\bfseries 64} no.~12, (2021) 120412},
  \href{http://arxiv.org/abs/2101.06869}{{\ttfamily arXiv:2101.06869
  [astro-ph.CO]}}.

\bibitem{Eardley:1973zuo}
D.~M. Eardley, D.~L. Lee, and A.~P. Lightman, ``{Gravitational-wave
  observations as a tool for testing relativistic gravity},''
  \href{http://dx.doi.org/10.1103/PhysRevD.8.3308}{{\em Phys. Rev. D}
  {\bfseries 8} (1973) 3308--3321}.

\bibitem{Eardley:1973br}
D.~M. Eardley, D.~L. Lee, A.~P. Lightman, R.~V. Wagoner, and C.~M. Will,
  ``{Gravitational-wave observations as a tool for testing relativistic
  gravity},'' \href{http://dx.doi.org/10.1103/PhysRevLett.30.884}{{\em Phys.
  Rev. Lett.} {\bfseries 30} (1973) 884--886}.

\bibitem{Chamberlin:2011ev}
S.~J. Chamberlin and X.~Siemens, ``{Stochastic backgrounds in alternative
  theories of gravity: overlap reduction functions for pulsar timing arrays},''
  \href{http://dx.doi.org/10.1103/PhysRevD.85.082001}{{\em Phys. Rev. D}
  {\bfseries 85} (2012) 082001},
  \href{http://arxiv.org/abs/1111.5661}{{\ttfamily arXiv:1111.5661
  [astro-ph.HE]}}.

\bibitem{Gair:2015hra}
J.~R. Gair, J.~D. Romano, and S.~R. Taylor, ``{Mapping gravitational-wave
  backgrounds of arbitrary polarisation using pulsar timing arrays},''
  \href{http://dx.doi.org/10.1103/PhysRevD.92.102003}{{\em Phys. Rev. D}
  {\bfseries 92} no.~10, (2015) 102003},
  \href{http://arxiv.org/abs/1506.08668}{{\ttfamily arXiv:1506.08668 [gr-qc]}}.

\bibitem{Cornish:2017oic}
N.~J. Cornish, L.~O'Beirne, S.~R. Taylor, and N.~Yunes, ``{Constraining
  alternative theories of gravity using pulsar timing arrays},''
  \href{http://dx.doi.org/10.1103/PhysRevLett.120.181101}{{\em Phys. Rev.
  Lett.} {\bfseries 120} no.~18, (2018) 181101},
  \href{http://arxiv.org/abs/1712.07132}{{\ttfamily arXiv:1712.07132 [gr-qc]}}.

\bibitem{Romano:2016dpx}
J.~D. Romano and N.~J. Cornish, ``{Detection methods for stochastic
  gravitational-wave backgrounds: a unified treatment},''
  \href{http://dx.doi.org/10.1007/s41114-017-0004-1}{{\em Living Rev. Rel.}
  {\bfseries 20} no.~1, (2017) 2},
  \href{http://arxiv.org/abs/1608.06889}{{\ttfamily arXiv:1608.06889 [gr-qc]}}.

\bibitem{NANOGrav:2021ini}
{\bfseries NANOGrav} Collaboration, Z.~Arzoumanian {\em et~al.}, ``{The
  NANOGrav 12.5-year data set: Search for Non-Einsteinian Polarization Modes in
  theGravitational-Wave Background},''
  \href{http://arxiv.org/abs/2109.14706}{{\ttfamily arXiv:2109.14706 [gr-qc]}}.

\bibitem{LIGOScientific:2021sio}
{\bfseries LIGO Scientific, VIRGO, KAGRA} Collaboration, R.~Abbott {\em
  et~al.}, ``{Tests of General Relativity with GWTC-3},''
  \href{http://arxiv.org/abs/2112.06861}{{\ttfamily arXiv:2112.06861 [gr-qc]}}.

\bibitem{Allen:1987bk}
B.~Allen, ``{The Stochastic Gravity Wave Background in Inflationary Universe
  Models},'' \href{http://dx.doi.org/10.1103/PhysRevD.37.2078}{{\em Phys. Rev.
  D} {\bfseries 37} (1988) 2078}.

\bibitem{Bartolo:2018qqn}
N.~Bartolo, V.~Domcke, D.~G. Figueroa, J.~Garc\'\i{}a-Bellido, M.~Peloso,
  M.~Pieroni, A.~Ricciardone, M.~Sakellariadou, L.~Sorbo, and G.~Tasinato,
  ``{Probing non-Gaussian Stochastic Gravitational Wave Backgrounds with
  LISA},'' \href{http://dx.doi.org/10.1088/1475-7516/2018/11/034}{{\em JCAP}
  {\bfseries 11} (2018) 034}, \href{http://arxiv.org/abs/1806.02819}{{\ttfamily
  arXiv:1806.02819 [astro-ph.CO]}}.

\bibitem{Bartolo:2018evs}
N.~Bartolo, V.~De~Luca, G.~Franciolini, A.~Lewis, M.~Peloso, and A.~Riotto,
  ``{Primordial Black Hole Dark Matter: LISA Serendipity},''
  \href{http://dx.doi.org/10.1103/PhysRevLett.122.211301}{{\em Phys. Rev.
  Lett.} {\bfseries 122} no.~21, (2019) 211301},
  \href{http://arxiv.org/abs/1810.12218}{{\ttfamily arXiv:1810.12218
  [astro-ph.CO]}}.

\bibitem{Bartolo:2018rku}
N.~Bartolo, V.~De~Luca, G.~Franciolini, M.~Peloso, D.~Racco, and A.~Riotto,
  ``{Testing primordial black holes as dark matter with LISA},''
  \href{http://dx.doi.org/10.1103/PhysRevD.99.103521}{{\em Phys. Rev. D}
  {\bfseries 99} no.~10, (2019) 103521},
  \href{http://arxiv.org/abs/1810.12224}{{\ttfamily arXiv:1810.12224
  [astro-ph.CO]}}.

\bibitem{Margalit:2020sxp}
A.~Margalit, C.~R. Contaldi, and M.~Pieroni, ``{Phase decoherence of
  gravitational wave backgrounds},''
  \href{http://dx.doi.org/10.1103/PhysRevD.102.083506}{{\em Phys. Rev. D}
  {\bfseries 102} no.~8, (2020) 083506},
  \href{http://arxiv.org/abs/2004.01727}{{\ttfamily arXiv:2004.01727
  [astro-ph.CO]}}.

\bibitem{Allen:1999xw}
B.~Allen, E.~E. Flanagan, and M.~A. Papa, ``{Is the squeezing of relic
  gravitational waves produced by inflation detectable?},''
  \href{http://dx.doi.org/10.1103/PhysRevD.61.024024}{{\em Phys. Rev. D}
  {\bfseries 61} (2000) 024024},
  \href{http://arxiv.org/abs/gr-qc/9906054}{{\ttfamily arXiv:gr-qc/9906054}}.

\bibitem{Bartolo:2019oiq}
N.~Bartolo, D.~Bertacca, S.~Matarrese, M.~Peloso, A.~Ricciardone, A.~Riotto,
  and G.~Tasinato, ``{Anisotropies and non-Gaussianity of the Cosmological
  Gravitational Wave Background},''
  \href{http://dx.doi.org/10.1103/PhysRevD.100.121501}{{\em Phys. Rev. D}
  {\bfseries 100} no.~12, (2019) 121501},
  \href{http://arxiv.org/abs/1908.00527}{{\ttfamily arXiv:1908.00527
  [astro-ph.CO]}}.

\bibitem{Bartolo:2019yeu}
N.~Bartolo, D.~Bertacca, S.~Matarrese, M.~Peloso, A.~Ricciardone, A.~Riotto,
  and G.~Tasinato, ``{Characterizing the cosmological gravitational wave
  background: Anisotropies and non-Gaussianity},''
  \href{http://dx.doi.org/10.1103/PhysRevD.102.023527}{{\em Phys. Rev. D}
  {\bfseries 102} no.~2, (2020) 023527},
  \href{http://arxiv.org/abs/1912.09433}{{\ttfamily arXiv:1912.09433
  [astro-ph.CO]}}.

\bibitem{Dimastrogiovanni:2019bfl}
E.~Dimastrogiovanni, M.~Fasiello, and G.~Tasinato, ``{Searching for Fossil
  Fields in the Gravity Sector},''
  \href{http://dx.doi.org/10.1103/PhysRevLett.124.061302}{{\em Phys. Rev.
  Lett.} {\bfseries 124} no.~6, (2020) 061302},
  \href{http://arxiv.org/abs/1906.07204}{{\ttfamily arXiv:1906.07204
  [astro-ph.CO]}}.

\bibitem{Powell:2019kid}
C.~Powell and G.~Tasinato, ``{Probing a stationary non-Gaussian background of
  stochastic gravitational waves with pulsar timing arrays},''
  \href{http://dx.doi.org/10.1088/1475-7516/2020/01/017}{{\em JCAP} {\bfseries
  01} (2020) 017}, \href{http://arxiv.org/abs/1910.04758}{{\ttfamily
  arXiv:1910.04758 [gr-qc]}}.

\bibitem{Green:2020whw}
D.~Green and R.~A. Porto, ``{Signals of a Quantum Universe},''
  \href{http://dx.doi.org/10.1103/PhysRevLett.124.251302}{{\em Phys. Rev.
  Lett.} {\bfseries 124} no.~25, (2020) 251302},
  \href{http://arxiv.org/abs/2001.09149}{{\ttfamily arXiv:2001.09149
  [hep-th]}}.

\bibitem{Adshead:2009bz}
P.~Adshead and E.~A. Lim, ``{3-pt Statistics of Cosmological Stochastic
  Gravitational Waves},''
  \href{http://dx.doi.org/10.1103/PhysRevD.82.024023}{{\em Phys. Rev. D}
  {\bfseries 82} (2010) 024023},
  \href{http://arxiv.org/abs/0912.1615}{{\ttfamily arXiv:0912.1615
  [astro-ph.CO]}}.

\bibitem{Maggiore:2018sht}
M.~Maggiore, {\em {Gravitational Waves. Vol. 2: Astrophysics and Cosmology}}.
\newblock Oxford University Press, 3, 2018.

\bibitem{Seto:2009ju}
N.~Seto, ``{Non-Gaussianity analysis of GW background made by short-duration
  burst signals},'' \href{http://dx.doi.org/10.1103/PhysRevD.80.043003}{{\em
  Phys. Rev. D} {\bfseries 80} (2009) 043003},
  \href{http://arxiv.org/abs/0908.0228}{{\ttfamily arXiv:0908.0228 [gr-qc]}}.

\bibitem{Anholm:2008wy}
M.~Anholm, S.~Ballmer, J.~D.~E. Creighton, L.~R. Price, and X.~Siemens,
  ``{Optimal strategies for gravitational wave stochastic background searches
  in pulsar timing data},''
  \href{http://dx.doi.org/10.1103/PhysRevD.79.084030}{{\em Phys. Rev. D}
  {\bfseries 79} (2009) 084030},
  \href{http://arxiv.org/abs/0809.0701}{{\ttfamily arXiv:0809.0701 [gr-qc]}}.

\bibitem{vanHaasteren:2008yh}
R.~van Haasteren, Y.~Levin, P.~McDonald, and T.~Lu, ``{On measuring the
  gravitational-wave background using Pulsar Timing Arrays},''
  \href{http://dx.doi.org/10.1111/j.1365-2966.2009.14590.x}{{\em Mon. Not. Roy.
  Astron. Soc.} {\bfseries 395} (2009) 1005},
  \href{http://arxiv.org/abs/0809.0791}{{\ttfamily arXiv:0809.0791
  [astro-ph]}}.

\bibitem{Lentati:2014hja}
L.~Lentati, M.~P. Hobson, and P.~Alexander, ``{Bayesian Estimation of
  Non-Gaussianity in Pulsar Timing Analysis},''
  \href{http://dx.doi.org/10.1093/mnras/stu1721}{{\em Mon. Not. Roy. Astron.
  Soc.} {\bfseries 444} no.~4, (2014) 3863--3878},
  \href{http://arxiv.org/abs/1405.2460}{{\ttfamily arXiv:1405.2460
  [astro-ph.IM]}}.

\bibitem{Demorest:2012bv}
P.~B. Demorest {\em et~al.}, ``{Limits on the Stochastic Gravitational Wave
  Background from the North American Nanohertz Observatory for Gravitational
  Waves},'' \href{http://dx.doi.org/10.1088/0004-637X/762/2/94}{{\em Astrophys.
  J.} {\bfseries 762} (2013) 94},
  \href{http://arxiv.org/abs/1201.6641}{{\ttfamily arXiv:1201.6641
  [astro-ph.CO]}}.

\bibitem{Chamberlin:2014ria}
S.~J. Chamberlin, J.~D.~E. Creighton, X.~Siemens, P.~Demorest, J.~Ellis, L.~R.
  Price, and J.~D. Romano, ``{Time-domain Implementation of the Optimal
  Cross-Correlation Statistic for Stochastic Gravitational-Wave Background
  Searches in Pulsar Timing Data},''
  \href{http://dx.doi.org/10.1103/PhysRevD.91.044048}{{\em Phys. Rev. D}
  {\bfseries 91} no.~4, (2015) 044048},
  \href{http://arxiv.org/abs/1410.8256}{{\ttfamily arXiv:1410.8256
  [astro-ph.IM]}}.

\bibitem{Vigeland:2018ipb}
S.~J. Vigeland, K.~Islo, S.~R. Taylor, and J.~A. Ellis, ``{Noise-marginalized
  optimal statistic: A robust hybrid frequentist-Bayesian statistic for the
  stochastic gravitational-wave background in pulsar timing arrays},''
  \href{http://dx.doi.org/10.1103/PhysRevD.98.044003}{{\em Phys. Rev. D}
  {\bfseries 98} (2018) 044003},
  \href{http://arxiv.org/abs/1805.12188}{{\ttfamily arXiv:1805.12188
  [astro-ph.IM]}}.

\bibitem{NANOGrav:2017wvv}
{\bfseries NANOGrav} Collaboration, Z.~Arzoumanian {\em et~al.}, ``{The
  NANOGrav 11-year Data Set: High-precision timing of 45 Millisecond
  Pulsars},'' \href{http://dx.doi.org/10.3847/1538-4365/aab5b0}{{\em Astrophys.
  J. Suppl.} {\bfseries 235} no.~2, (2018) 37},
  \href{http://arxiv.org/abs/1801.01837}{{\ttfamily arXiv:1801.01837
  [astro-ph.HE]}}.

\bibitem{Jenet:2014bea}
F.~A. Jenet and J.~D. Romano, ``{Understanding the gravitational-wave Hellings
  and Downs curve for pulsar timing arrays in terms of sound and
  electromagnetic waves},'' \href{http://dx.doi.org/10.1119/1.4916358}{{\em Am.
  J. Phys.} {\bfseries 83} (2015) 635},
  \href{http://arxiv.org/abs/1412.1142}{{\ttfamily arXiv:1412.1142 [gr-qc]}}.

\bibitem{Thrane:2013oya}
E.~Thrane and J.~D. Romano, ``{Sensitivity curves for searches for
  gravitational-wave backgrounds},''
  \href{http://dx.doi.org/10.1103/PhysRevD.88.124032}{{\em Phys. Rev. D}
  {\bfseries 88} no.~12, (2013) 124032},
  \href{http://arxiv.org/abs/1310.5300}{{\ttfamily arXiv:1310.5300
  [astro-ph.IM]}}.

\bibitem{Maggiore:2007ulw}
M.~Maggiore, {\em {Gravitational Waves. Vol. 1: Theory and Experiments}}.
\newblock Oxford Master Series in Physics. Oxford University Press, 2007.


\end{thebibliography}
\end{document}